\documentclass[a4paper,11pt]{article}

\usepackage{jheppub}
\usepackage[table,xcdraw]{xcolor}
\usepackage[utf8]{inputenc}
\usepackage{float}
\usepackage{amsmath}
\usepackage{graphicx}
\usepackage{multirow}
\usepackage{booktabs}
\usepackage{subcaption}
\usepackage{tikz}
\usepackage{xfp}
\usepackage{diagbox}
\usepackage[flushleft]{threeparttable}

\DeclareMathOperator{\diag}{diag}

\newcommand{\orcid}[1]{\href{https://orcid.org/#1}{#1}}

\definecolor{mygreen}{rgb}{0.6,1,0.6}        
\definecolor{myyellow}{rgb}{1,1,0.5}           
\definecolor{myorange}{rgb}{1,0.55,0.0}         
\definecolor{darkred}{rgb}{0.85,0,0}  
\definecolor{LightRed}{rgb}{1, 0.8, 0.8} 

\begin{document}

\title{The Future of Lepton Flavor}

\author[1]{Peter B.~Denton\note{\orcid{0000-0002-5209-872X}},}
\author[2]{Julia Gehrlein\note{\orcid{0000-0002-1235-0505}},}
\author[2]{and Henry Truelson\note{\orcid{0009-0002-4320-4049}}}

\affiliation[1]{High Energy Theory Group, Physics Department, Brookhaven National Laboratory, Upton, NY 11973, USA}
\affiliation[2]{Physics Department, Colorado State University, Fort Collins,   CO 80523, USA}

\emailAdd{pdenton@bnl.gov}
\emailAdd{julia.gehrlein@colostate.edu}
\emailAdd{henry.truelson@colostate.edu}
\makeatletter
\hypersetup{colorlinks=true,allcolors=[rgb]{1,0.56,0}}
\makeatother

\abstract{
The flavor puzzle remains one of the biggest open questions in particle theory to date and upcoming results from neutrino experiments will have a large impact on its potential solution in the future.
While some classes of leptonic flavor models are difficult to constrain with current data, this will change in the coming years as several yet unknown quantities, like the neutrino mass ordering and the octant of $\theta_{23}$, will be determined, and the CP-violating quantity $\delta$ will be measured with some precision.
In addition, significant improvements in the precision of the other oscillation parameters, notably $\theta_{12}$, is also expected to impact our understanding of flavor.
Together with anticipated improvements on the absolute neutrino mass scale determination from the combination of cosmological data sets or beta decay endpoint spectrum measurements, upcoming experiments will lead to a refined picture of our understanding of flavor in the lepton sector.
In this paper, we show exactly how flavor model predictions relate to expected measurements.
Five popular classes of leptonic flavor model predictions are considered: mass sum rules, one and two texture-zeros for both Dirac and Majorana neutrinos, charged lepton corrections, modular symmetries, and constrained sequential dominance.
We discuss correlations, degeneracies, and discrimination capabilities in the context of the expected measurements from upcoming experiments. We also highlight how different flavor model predictions can be differentiated and  the roles each upcoming observable has on flavor model predictions.
We anticipate that the precision targeted in future measurements will be sufficient to dramatically reduce the number of viable leptonic flavor models, will allow us to disentangle them, and could hopefully begin to shed light on the answer to the flavor puzzle.}

\maketitle

\section{Introduction}
The Standard Model (SM) of particle physics encodes all thus far observed particles and interactions with just a small number of free parameters. The majority of these free parameters is related to the flavor sector. These parameters do not seem to be obviously related to each other or other parameters in the model and the charged fermion masses span six orders of magnitude \cite{ParticleDataGroup:2024cfk}. Furthermore, the discovery of neutrino oscillations \cite{SNO:2002tuh,Super-Kamiokande:1998kpq} further complicates the flavor puzzle as it introduced at least seven new parameters to the model and seemingly a new scale, even more removed from the electroweak scale. As neutrino oscillation experiments have steadily improved the precision of the remaining parameters, it has became clear that the leptonic mixing matrix features much more mixing than its analog in the quark sector \cite{ParticleDataGroup:2024cfk}. Additionally, the neutrino masses are much smaller than the lightest charged fermion, further complicating a straightforward understanding of the origin of flavor.

A huge swath of symmetry-related ideas that aim to provide a rationale behind the observed leptonic mixing pattern and the smallness of neutrino masses
have been proposed over the years, see e.g.~\cite{King:2013eh,King:2014nza,Xing:2015fdg,Petcov:2017ggy,Xing:2020ijf,Gehrlein:2022nss,Chauhan:2023faf,Altmannshofer:2024hmr} for several reviews, but no clear contender has emerged so far.
A shared characteristic of the vast majority of flavor models is that any new particles and physical scales are likely beyond the reach of high-energy colliders as the scale of flavor symmetry breaking typically resides at a very high scale, depending on the underlying model.
Commonly discussed scales relate to the seesaw mechanism or grand unification around $10^{15}$ GeV.\footnote{There are some notable exceptions to this statement and models with a low-scale flavor exist, see for example \cite{Grinstein:2010ve,Alonso:2013nca,
Alonso:2016onw,FernandezNavarro:2023rhv,Ibarra:2017tju,Kitabayashi:2019qvi,Hong:2019bki,Arkani-Hamed:2026wwy}.
These model predictions can additionally be tested at colliders and lepton flavor experiments.} Therefore, precisely testing their low-energy predictions likely provides the only way to test and disentangle flavor models.
Different classes of models tend to behave quite differently both theoretically and phenomenologically, and relating them proves challenging.
Therefore, many studies focus on one class of models only, see for example \cite{Barry:2010yk,King:2013psa,Bergstrom:2014owa,Ballett:2013wya,Ballett:2014dua,Girardi:2015vha,Agarwalla:2017wct,Blennow:2020snb,Blennow:2020ncm}.
Nonetheless, there have been a number of studies attempting to make  comparisons between different classes.
In \cite{Gehrlein:2022nss} an overview of predictions for the oscillation parameters from various classes of flavor models has been provided.
Reference \cite{Denton:2023hkx} considered phenomenological model predictions  and cast the results in the context of neutrinoless double beta decay ($0\nu\beta\beta$) and absolute neutrino scale measurements, particularly paying attention to the funnel in the normal mass ordering  where the $0\nu\beta\beta$ rate can be arbitrarily small (see also \cite{Plentinger:2006nb,Jenkins:2008ex} for earlier general studies before $\theta_{13}$ was measured). 

As measurements of the quark and charged lepton sectors are already highly precise, most of the remaining freedom for theoretical predictions lies in the neutrino sector \cite{Capozzi:2025ovi,Esteban:2026phq}.
The seven guaranteed low-energy fundamental parameters to measure and predict are three masses, three mixing angles, and one complex phase (in the usual parameterization).
If neutrinos are Majorana, then two additional phases are physical, although they cannot be easily determined even if we confirm that neutrinos are Majorana particles.
While we have a good understanding of many of the neutrino parameters, the experimental program is poised to make a giant leap forward in the coming years \cite{Denton:2022een,Denton:2025jkt}.
Four major unknowns remain in the neutrino sector: the sign of $\Delta m^2_{31}$ known as the atmospheric mass ordering (MO), the octant of $\theta_{23}$, the value of the  phase $\delta$, and the absolute neutrino mass scale.
The MO will be well determined in the coming years by DUNE \cite{DUNE:2020ypp}, JUNO \cite{JUNO:2021vlw}, with atmospheric neutrinos at Hyper-Kamiokande (HK), IceCube, and KM3NeT \cite{Hyper-Kamiokande:2018ofw,IceCube:2019dyb,KM3NeT:2021ozk}, and the combination of data sets \cite{Nunokawa:2005nx,IceCube-Gen2:2019fet,Parke:2024xre}.
The octant of $\theta_{23}$ and CP violation, parameterized by $\delta$, are somewhat challenging, but long-baseline accelerator appearance measurements at DUNE and HK will provide good information on both.
Finally, the absolute mass scale requires non-oscillation data.
The strongest sensitivity comes from combining cosmological data sets, but current data exhibits tensions  that may be related to neutrino masses and are not yet understood \cite{Planck:2013nga,Couchot:2017pvz,eBOSS:2020yzd,Abdalla:2022yfr,Craig:2024tky,Wang:2024hen,Allali:2024aiv,Yadav:2024duq,Green:2024xbb,Elbers:2024sha,Naredo-Tuero:2024sgf,Jiang:2024viw,Bertolez-Martinez:2024wez,Lynch:2025ine,CosmoVerseNetwork:2025alb,Sailer:2025lxj,Cozzumbo:2025ewt}; until these issues are fully sorted, drawing strong conclusions about neutrino masses from cosmology is suspect. Complementary information exists from 
the KATRIN experiment which aims to make a straightforward measurement of the absolute mass scale by looking for a modification of the endpoint of the spectrum of beta decay and has limits at the $\sim0.5$~eV scale \cite{KATRIN:2024cdt}. 
Upcoming laboratory based experiments like the ECHo experiment \cite{Gastaldo:2017edk}, Project 8 \cite{Project8:2017nal}, potential improvements of tritium based experiments \cite{electron}, and the PTOLEMY experiment \cite{PTOLEMY:2019hkd} as well as cosmological experiments \cite{Font-Ribera:2013rwa,SimonsObservatory:2018koc,CMB-S4:2016ple,Brinckmann:2018owf}  will be sensitive down to neutrino masses in the $\mathcal{O}(10~\text{meV})$ region and hopefully will provide not only a clear discovery of the absolute neutrino mass scale, but a determination with some precision.
Beyond these open questions, we also anticipate significant, pushing an order of magnitude, improvement in the precision on many of the other parameters, notably $\theta_{12}$, the solar mixing angle which plays a key role in many flavor models, $\Delta m^2_{21}$ the solar mass splitting, and $\Delta m^2_{31}$ the atmospheric mass splitting; the reactor mixing angle $\theta_{13}$ is not expected to significantly improve, however its current precision is already driving many of the constraints on flavor models thus any progress there is also useful.

Given the combination of this promising experimental landscape and the challenging theoretical picture, we have investigated how our knowledge of the flavor model landscape is expected to evolve 
in the coming years. 
We have highlighted the key observables for different classes  of models and identified where correlated predictions can be used to differentiate among otherwise similar model predictions.
This provides a guidebook to understand the impact measurements of the remaining oscillation parameters and the absolute neutrino mass scale will have on our understanding of the flavor puzzle.

In this study, we will follow and expand upon the classification of model predictions presented in \cite{Denton:2023hkx}. We are focused on classes of models which are most predictive, in the sense that in addition to predicting neutrino parameters, they also provide correlations between them, opening the possibility to probe these models at a variety of experiments. However, we will remain agnostic about the underlying model and particle content, but instead classify the models in different classes based on their low-energy predictions. Therefore, we do not consider renormalization group running effects to the predictions from these flavor models, which depend on the exact particle content and the scale at which these predictions arise.  

The flavor model prediction classes we consider are mass sum rules (section~\ref{sec:sum rules}), texture-zeros with one or two zeros and for either Majorana or Dirac neutrinos (section~\ref{sec:texture zeros}), charged lepton corrections (section~\ref{sec:clc}), modular symmetries (section~\ref{sec:modular}), and constrained sequential dominance (section~\ref{sec:sequential}).
In each section of this paper, we will review the model, perform calculations of the allowed regions of the relevant parameters highlighting degeneracies, discrimination capabilities, and validity with the current data.
Moreover, going beyond only the predictions for $0\nu\beta\beta$, here we investigate the relationships among the model predictions and all oscillation parameters and absolute mass scale results.
This paints a clear picture
of which flavor models can be easily discriminated from others with upcoming measurements, and which will remain degenerate with others.
We focus on the above classes of models and do not include others such as generalized CP \cite{Wolfenstein:1981rk,Kayser:1984ge,Bilenky:1984fg,Branco:1986gr,Feruglio:2012cw,Holthausen:2012dk,Ding:2013nsa,Ding:2014hva,King:2014rwa,Ding:2014ssa,Hagedorn:2014wha,Ding:2014ora,Ding:2015rwa} which predicts the phases to have CP conserving values as these predictions are straightforward, anarchy \cite{Hall:1999sn,Haba:2000be,Jenkins:2008ms,deGouvea:2012ac} as the predictions on the fundamental parameters do not evolve with further measurements, Froggatt-Nielsen type models \cite{Froggatt:1978nt,Leurer:1992wg,Leurer:1993gy} as they feature a large number of free parameters, most of their values are driven by the measurements  in the charged lepton sector such that more precise measurements of the neutrino parameters will have less significant effects see also \cite{Cornella:2024jaw,Ibe:2024cvi}, or other models that don't fit into straightforward classifications such as models with gauged  SM  flavor symmetries \cite{King:2003rf,Grinstein:2010ve,DAgnolo:2012ulg,Alonso:2013nca,Alonso:2016onw,Greljo:2024zrj,Greljo:2023bix} or extended gauge symmetries for example \cite{Li:1981nk,Bordone:2017bld,FernandezNavarro:2023rhv,Capdevila:2024gki,Fuentes-Martin:2024fpx,Covone:2024elw,Fabri:2025fsc,Davighi:2025cqx}, models which focus on
lowering the scale of flavor via the introduction of chains of several new fermions or multi-scales
\cite{Ibarra:2017tju,Kitabayashi:2019qvi,Hong:2019bki,Arkani-Hamed:2026wwy,Berezhiani:1992pj,Barbieri:1994cx,Panico:2016ull}, and extra dimensions \cite{Grossman:1999ra,Gherghetta:2000qt,Huber:2000ie,Fuentes-Martin:2022xnb}, see also \cite{Chen:2005mz,Fong:2011xh}, or partial compositeness \cite{Kaplan:1991dc}.
Before we discuss the model classes, we will first present our methodology in the next section, and we end the paper with a discussion of our results in section~\ref{sec:discussion} and our conclusions in section~\ref{sec:conclusions}.

\section{Methodology}
\label{sec:methodology}
In the following sections, we
will derive the predictions of  various  classes of flavor models, compare them to the current neutrino data, and answer the question if upcoming  precision neutrino experiment can  distinguish these model predictions from one another. 

Throughout this paper we use the standard parametrization \cite{ParticleDataGroup:2024cfk,Denton:2020igp} of the PMNS matrix $U$ \cite{Pontecorvo:1957cp,Maki:1962mu}  with  three flavor mixing angles $\theta_{12},\theta_{13}, \theta_{23}$ and one Dirac CP-violating phase $\delta$ as
\begin{equation}
   U = U_{PMNS} \equiv \begin{pmatrix}
    c_{12} c_{13}& c_{13}s_{12}& e^{-i\delta}s_{13}\\
    -c_{23} s_{12} - e^{i\delta}c_{12} s_{13} s_{23} & c_{12} c_{23} - e^{i\delta}s_{12} s_{13} s_{23}& c_{13} s_{23}\\
    -e^{-i\delta}c_{12} c_{23} s_{13}+ s_{12} s_{23} & -e^{i\delta}c_{23} s_{12} s_{13} - c_{12} s_{23} & c_{13} c_{23}
\end{pmatrix}\,,
\label{eq:PMNSParametrization}
\end{equation}
where $c_{ij} = \cos\theta_{ij}$ and $s_{ij} = \sin\theta_{ij}$. 
To extract the mixing parameters from a unitary matrix we use that 

\begin{equation}
s^2_{13} = |(U_{PMNS})_{e3}|^2,\;\; s_{12}^2 = \frac{|(U_{PMNS})_{e2}|^2}{1-|(U_{PMNS})_{e3}|^2},\;\; s_{23}^2 = \frac{|(U_{PMNS})_{\mu3}|^2}{1-|(U_{PMNS})_{e3}|^2}\,,
\label{eq:PMNSExtraction}
\end{equation}
 while the expression for $\delta$ is more involved, but can be determined as the phase of $(U_{PMNS})_{e3}$ subject to the correct rephasing conditions, and sign information from the Jarlskog invariant \cite{Jarlskog:1985ht}.
 
To find the model prediction parameter space compatible with current neutrino data we construct a $\chi^2$ function using the NuFit-v6 results from \cite{Esteban:2024eli} and recent JUNO results  from \cite{JUNO:2025gmd}  for the mixing parameters in the PMNS matrix \cite{Pontecorvo:1957cp,Maki:1962mu}, see table~\ref{tab:OscillationParameters}. 
For our analyses, we assume  no prior on the CP-violating phase $\delta$ as current experiments only provide a weak hint for a preferred value \cite{NOvA:2021nfi,T2K:2025yoy} which also slightly disagrees between T2K and NOvA  \cite{T2K:2025wet}. Note however, that other studies of these model predictions in the literature  do include a prior on $\delta$, differences in our results compared to literature results are largely attributed to this difference in approach. 
Currently, the MO is not measured yet. 
The neutrino mass could either be normal ordered (NO) with $m_1<m_2<m_3$ or inverted ordered (IO) with $m_3<m_1<m_2$ where the mass eigenstates are defined as $(U_{PMNS})_{e1}>(U_{PMNS})_{e2}>(U_{PMNS})_{e3}$. From global neutrino data, there is only a weak preference for the NO, therefore, we do not assume a prior on the MO. 
If the model predictions include the neutrino masses, we phrase them in terms of the lightest neutrino mass $m_\ell$, which is $m_1$ for NO and $m_3$ for IO.\footnote{
We don't take correlations in the experimental results into account; there are no significant correlations in the parameters best measured by reactor neutrinos (solar and $\theta_{13}$); while there are some correlations among $\theta_{23}$, $\delta$, and the MO, as we do not consider any information on $\delta$, the MO, or the octant, these are not relevant for our numerical analysis.}
Finally, as we discuss in sec.~\ref{sec:model_overview}, four classes of flavor models analyzed here make predictions on the absolute neutrino mass scale. In our analysis, we choose a conservative approach to not include a $\chi^2$ contribution from mass measurements but instead focus on our results in the region up to $m_\ell <1$ eV.
We also show in table \ref{tab:OscillationParameters} the constraints on the absolute mass scale; these are not included in the statistical fits due to the complexity of the cosmology data, but we do refer to these results when discussing the validity of model predictions.
Precision measurements of the absolute mass scale will be challenging, especially if one considers the constraint from cosmological data indicating a low mass scale, and the fact that cosmology measures $\sum m_i$ which translates to a rather large error on the lightest neutrino mass.
To identify the best fit point and the predicted region we scan through the model prediction parameter space, in $\log(m_\ell)$ and linear in the mixing angles.

\begin{table}
  \centering
  \caption{Compiled results for the  neutrino oscillation parameters from \cite{Esteban:2024eli,JUNO:2025gmd}. We include here their best fit values with $1\sigma$ uncertainties and their respective $3\sigma$ ranges. We assume no prior on $\delta$ for our analyses.
  We show the absolute mass scale constraints from 
  \cite{KATRIN:2024cdt, DESI:2024hhd}
  (cosmological data only provides an upper limit for the NO; it nominally disfavors the IO) but do not use this data in statistical analyses.}
  \begin{threeparttable}
    \renewcommand{\arraystretch}{1.3}
    \begin{tabular}{l cccc r}
    \toprule
    \toprule
    Parameter & \multicolumn{2}{c}{Best-fit $\pm$ $1\sigma$} & \multicolumn{2}{c}{$3\sigma$ Range} & Ref.\\
    \cmidrule(lr){2-3} \cmidrule(lr){4-5}
    & NO & IO & NO & IO & \\
    \midrule
    $\frac{\Delta m_{31}^2}{10^{-3}\,\text{eV}^2}$ & $2.534^{+0.025}_{-0.023}$ & --- & $2.463 \to 2.606$ & --- & \multirow{6}{*}{\cite{Esteban:2024eli}} \\
    $\frac{\Delta m_{32}^2}{10^{-3}\,\text{eV}^2}$ & --- & $-2.510^{+0.024}_{-0.025}$ & --- & $-2.584 \to -2.438$ & \\
    $\sin^2\theta_{13}$ & $0.02195^{+0.00054}_{-0.00058}$ & $0.02195^{+0.00054}_{-0.00058}$ & $0.02023 \to 0.02376$ & $0.02053 \to 0.02397$ & \\
    $\theta_{13}(^\circ)$ & $8.52^{+0.11}_{-0.11}$ & $8.52^{+0.11}_{-0.11}$ & $8.18 \to 8.87$ & $8.24 \to 8.91$ & \\
    $\sin^2\theta_{23}$ & $0.561^{+0.012}_{-0.015}$ & $0.561^{+0.012}_{-0.015}$ & $0.430 \to 0.596$ & $0.437 \to 0.597$ & \\
    $\theta_{23}(^\circ)$ & $48.5^{+0.7}_{-0.9}$ & $48.6^{+0.7}_{-0.9}$ & $41.0 \to 50.5$ & $41.4 \to 50.6$ & \\
    \midrule
    $\frac{\Delta m_{21}^2}{10^{-5}\text{eV}^2}$ & \multicolumn{2}{c}{$7.50^{+0.12}_{-0.12}$} & \multicolumn{2}{c}{$7.14 \to 7.86$} & \multirow{3}{*}{\cite{JUNO:2025gmd}} \\
    $\sin^2\theta_{12}$ & \multicolumn{2}{c}{$0.3092^{+0.0087}_{-0.0087}$} & \multicolumn{2}{c}{$0.2834 \to 0.3356$} & \\
    $\theta_{12}(^\circ)$ & \multicolumn{2}{c}{$33.78^{+0.54}_{-0.54}$} & \multicolumn{2}{c}{$32.16 \to 35.40$} & \\
    \midrule
    $m_\nu^2$ \scriptsize{(KATRIN)} & \multicolumn{2}{c}{---} & \multicolumn{2}{c}{$m_\nu^2 < 0.28$ eV$^2$} & \cite{KATRIN:2024cdt} \\
     $m_1$ \scriptsize{(DESI)} & \multicolumn{2}{c}{---} & \multicolumn{2}{c}{$m_1 < 0.025$ eV} & \cite{DESI:2024hhd} \\
    \bottomrule
    \bottomrule
    \end{tabular}
  \end{threeparttable}
  \label{tab:OscillationParameters}
\end{table}

In the following, we declare a model prediction invalid if its best fit point has a $\chi^2_{bf}$ per degree of freedom much larger than one  which ensures that the predictions can accommodate the current knowledge on neutrino data. We then compare the predictions for the neutrino parameters to determine the overlap between the two model prediction parameter spaces. More aligned parameter spaces for the neutrino parameters make two model predictions more difficult to distinguish by  measurements, while less compatible model predictions can more easily allow us to differentiate between them using upcoming measurements. 
To quantify the amount of alignment between two model prediction parameter spaces,  we introduce an asymmetric integral measurement.   For an assumed ``true'' reference model $A$ and ``test'' model $B$ making predictions on the parameters $\theta_i$, we compute the expectation value of the relative likelihood ratio of the predictions of model $B$ to the predictions of model $A$ as
\begin{equation}
    R(A||B) = \int \frac{\mathcal{L}(D|\theta_i, B)}{\mathcal{L}(D | \theta_i, A)}P(\theta_i | D, A) d\theta_i.
    \label{eq:IntegralOverlap}
\end{equation}
with the probability distribution for the model prediction parameter space $A$ as $P(\theta_i | D, A)$ for parameters $\theta_i$ and $D$ stands for the data. In the case that one parameter is not explicitly involved in an individual model prediction, integration is still carried out over that parameter within its defined experimental bounds. This happens only for one model prediction,  that being $m_{ee}=0$ in the texture-zero class discussed in sec.~\ref{sec:texture zeros}. 
We assume the standard Gaussian likelihood ratio
\begin{equation}
\mathcal{L}(D | \theta,H) = e^{-\frac{1}{2}\Delta \chi^2_{H}}\,,
\end{equation}
with $\Delta \chi^2_H = \chi^2_H-\chi^2_{min,H}$ for the hypothesis $H$ and define the probability distribution as
\begin{equation}
P(\theta|D,H) = \frac{1}{N_H} \mathcal{L}(D|\theta,H)\,,
\end{equation}
where $N_H$ is the normalization factor to ensure the probability is normalized.

The result $R(A||B) \rightarrow 0$ implies the model prediction parameter spaces are completely disjoint in their predictions
and are more easily distinguished by a measurement. The result $R(A||B) \rightarrow 1$
implies that the predictions of model $B$ are degenerate with the predictions of model $A$ across the credible volume of $A$, and therefore if $A$ is the ``true'' model, the observed data will be equally compatible with model $B$, rendering the two models indistinguishable by their predictions. Note that $R$ is antisymmetric, in general $R(A||B)\not = R(B||A)$, and $R(A||A) = 1$ by construction.
The integration region is the entire parameter space $\theta_i \in \mathcal{M}$. In practice, the integral is evaluated using Metropolis-Hastings Monte Carlo importance sampling, where the points are distributed according to the probability distribution of the ``true'' model $A$.

To compare the model predictions to expected results from upcoming experiments, we use the forecasted sensitivities of 6 years of JUNO on the solar mixing parameters \cite{JUNO:2022mxj}, the expected sensitivities of DUNE and HK for $\theta_{23}$ and $\delta $ \cite{DUNE:2020ypp,Hyper-Kamiokande:2018ofw} using
600 kt-MW-yrs for DUNE and 10 years of HK \cite{Hyper-Kamiokande:2025fci}. For the precision on $\theta_{23}$ we assume  that the  constraint on $\theta_{13}$ from reactor experiments \cite{DayaBay:2022orm}   has been included.
As we are interested in the question if future experiments can distinguish between two model predictions, we assume that one best-fit point is realized in nature and can be measured with the expected precision from these experiments.

Figure \ref{fig:Model_Overview} gives an overview of the five classes of neutrino flavor models we consider in this work and the relevant mixing parameters that each model class predicts. As stated, we do not explicitly analyze the predictions on the Majorana phases. We find two 
main categories of model classes: those that make predictions on the absolute neutrino mass scale $m_\ell$, and those that do not.

\subsection{Classes with neutrino mass scale predictions}
Several classes of models  make predictions for the absolute neutrino mass scale, cf.~the model predictions described in secs.~\ref{sec:sum rules}, \ref{sec:texture zeros}, \ref{sec:modular}.\footnote{Model predictions based on constrained sequential dominance in sec.~\ref{sec:sequential} predict $m_1=0$ but not a relation of the form in eq.~\eqref{eq:GeneralSumRule}. }  In these cases, a relationship between the three physical neutrino mass eigenstates of the general form arises
\begin{equation}
    C_1 e^{i\chi_1} \bar{m}_1^d + C_2 e^{i\chi_2}\bar{m}_2^d + m_3^d = 0\,,
    \label{eq:GeneralSumRule}
\end{equation}
where $C_1,C_2 \in \mathbb{R^+}$, $\chi_1,\chi_2\in\left[0,2\pi\right)$, and $d$ are defined by the underlying model, and $\bar{m}_1$, $\bar{m}_2$, and $m_3$ are the physical eigenvalues of the neutrino mass matrix where we already have rotated unphysical phases away. For Majorana neutrinos, we include the relative Majorana phases $\alpha,\beta$ such that $\bar{m}_1 = m_1 e^{i\alpha}$, and $\bar{m}_2 = m_2 e^{i\beta}$, whereas for Dirac neutrinos $\bar{m}_1 = m_1$ and $\bar{m}_2 = m_2$. We have chosen our expression such that for $m_3$ the coefficient is normalized to one with no phase.

\section{Overview of models}
\label{sec:model_overview}

\begin{figure}
\centering
    \includegraphics[width=.5\textwidth]{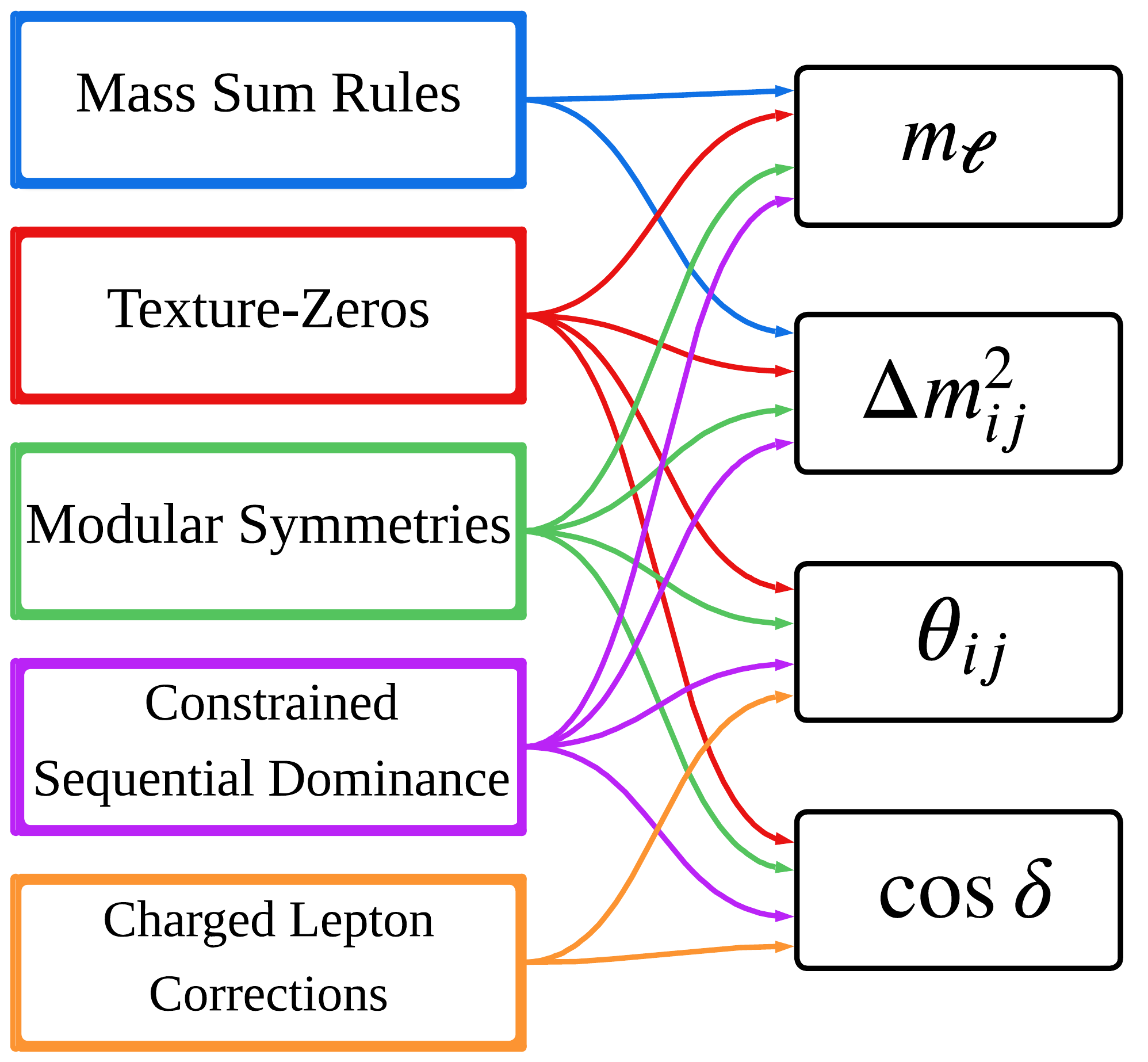}
    \caption{The five model classes analyzed in this work and the neutrino observables they  predict.}\label{fig:Model_Overview}
\end{figure}

Depending on the class of model, the coefficients $C_i$ are either constant, as in the case of neutrino mass sum rules discussed in sec.~\ref{sec:sum rules}, functions of the PMNS matrix elements, as is the case for texture-zeros discussed in sec.~\ref{sec:texture zeros}, or they depend on underlying model parameters which simultaneously determine the mixing parameters, as is the case for modular symmetries discussed in sec.~\ref{sec:modular}.

\begin{figure}
\centering
    \includegraphics[width=.9\textwidth]{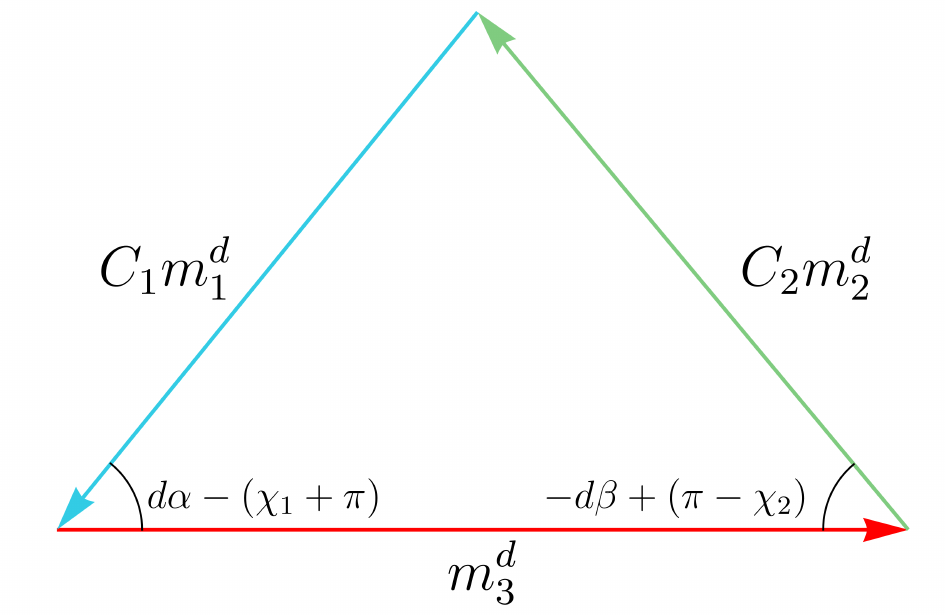}
    \caption{Geometric representation of eq.~\eqref{eq:GeneralSumRule} for Majorana neutrinos. Shown is an example of a mass sum rule with $C_1 = C_2 = d= 1$. This triangle represents a lightest mass prediction around $m_1 \sim 60$ meV for the mass splittings  at their best fit values.}
    \label{fig:SumRuleTriangle}
\end{figure}

Sum rules of this form  define two complex equations leading to correlations between parameters.
They can be interpreted as triangles in the complex plane, as illustrated in fig.~\ref{fig:SumRuleTriangle}, and hence the valid parameter space  is  constrained by the formation of a triangle. This means that 
the triangle inequality has to be fulfilled, which for  $|d|\geq 1$ leads to the constraints\footnote{For the case of $|d|<1$, the resulting expressions are more involved. We use eq.~\eqref{Eq:LawofCosines1} and~\eqref{Eq:LawofCosines2}  to define the valid model prediction parameter space.} 
\begin{equation}
 \frac{1}{C_1}\left|C_2 m_2^d - m_3^d\right| \leq m_1^d \leq \frac{1}{C_1}\left(C_2 m_2^d +m_3^d\right)\,,
     \label{eq:TriangleInequalityNO}
\end{equation}
\begin{equation}
     \left| C_1 m_1^d -C_2 m_2^d\right| \leq m_3^d \leq C_1 m_1^d+C_2 m_2^d\,.
     \label{eq:TriangleInequalityIO}
\end{equation}
The phases $\chi_1$ and $\chi_2$ are only relevant for predictions on the Majorana phases which can be extracted using the law of cosines
\begin{equation}
    \cos\left(d\alpha-\chi_1\right)= \frac{C_2^2m_2^{2d}-C_1^2m_1^{2d}-m_3^{2d}}{2C_1 m_1^dm_3^d}\,,
    \label{Eq:LawofCosines1}
\end{equation}
\begin{equation}
    \cos\left(d\beta+\chi_2\right)= \frac{C_1^2m_1^{2d}-C_2^2m_2^{2d}-m_3^{2d}}{2C_2 m_2^dm_3^d}\,.
    \label{Eq:LawofCosines2}
\end{equation}
In the case of Dirac neutrinos, the phases $\alpha,\beta$ are unphysical, and to satisfy eq.~\eqref{eq:GeneralSumRule}  either $\chi_1 = 0,\pi$ and $\chi_2 = 0,\pi$ or both $\chi_1,\chi_2 \not = 0,\pi$ are required. The model predictions on the absolute mass scale therefore depend only on $\left\{C_1, C_2, d\right\}$.
If $C_1$ and $C_2$ are of similar order of magnitude, which is the case in several model predictions, the lower bounds on the lightest mass in IO is typically smaller than in NO as $m_1$ and $m_2$ are closer together.

From eq.~\eqref{eq:GeneralSumRule}, one can see that model predictions containing the Majorana phases do not in general predict an upper limit on the mass scale, which is the case for most model predictions we study. However, when the angles of the triangle can no longer vary freely, as is the case with model predictions for Dirac neutrinos we study, the side lengths cannot be made arbitrarily large and hence they do necessarily impose an upper bound on the predictions of $m_\ell$.  

\subsection{Classes without neutrino mass scale predictions}
We also analyze classes of models which do not predict the absolute neutrino mass scale but only predict the PMNS mixing matrix. These are the model predictions with non-diagonal charged lepton mixing matrices discussed in sec.~\ref{sec:clc} and those based on constrained sequential dominance where one column of the mixing matrix is fixed as discussed in sec.~\ref{sec:sequential}. 

The model predictions discussed in secs.~\ref{sec:clc}, \ref{sec:modular}, \ref{sec:sequential} feature a PMNS matrix which can be written as the product of two mixing matrices, one of them features a zero or small $\theta_{13}$ angle. From the real and imaginary parts of the $e$-3 element of the PMNS matrix with the standard rephasing, we can extract the prediction for $\delta$.
Specifically we note that all of these model predictions are on $\cos\delta$, while the functional form of the predictions of the mixing angles vary.
For our numerical results, we use the parametrization-independent methods of extracting the mixing parameters in eq.~\eqref{eq:PMNSExtraction} to get the predictions for each parameter. 

\section{Neutrino mass sum rules}
\label{sec:sum rules}
Neutrino mass sum rules feature a relation between the three light neutrino masses of the form in eq.~\eqref{eq:GeneralSumRule}, where $C_i,\chi_i$ and $d$ are constants predicted in the underlying model \cite{Barry:2010yk,King:2013psa}. However, there is no one-to-one correspondence between the parameters of a mass sum rule and the details of the underlying model; the only requirement  for the appearance of a mass sum rule is that the three neutrino masses can be parametrized by only two parameters, hence leading to a relation between them \cite{Gehrlein:2017ryu}.
Neutrino mass sum rules arise both in models with Dirac neutrinos or models with Majorana neutrinos.
They make predictions on $m_\ell$ and the two mass splittings. Additionally, in the case of Majorana neutrinos, the Majorana phases can be predicted. 
Several mass sum rules can only be fulfilled in one MO, as we will see.

Up to now 12 mass sum rules have been identified in the literature which appear in over 60 different models based on discrete flavor symmetries \cite{Barry:2010yk,Bazzocchi:2009da,Ding:2010pc,Ma:2005sha,Ma:2006wm,Kang:2015xfa,Honda:2008rs,Brahmachari:2008fn,Altarelli:2005yx,Chen:2009um,Chen:2009gy,Cooper:2012bd,Altarelli:2009kr,Altarelli:2008bg,Hirsch:2008rp,Bazzocchi:2009pv,Everett:2008et,Boucenna:2012qb,Mohapatra:2012tb,Altarelli:2005yp,Altarelli:2006kg,Ma:2006vq,Bazzocchi:2007na,Bazzocchi:2007au,Lin:2008aj,Ma:2009wi,Ciafaloni:2009qs,Bazzocchi:2008ej,Feruglio:2013hia,Chen:2007afa,Ding:2008rj,Chen:2009gf,Feruglio:2007uu,Merlo:2011hw,Luhn:2012bc,Fukuyama:2010mz,Ding:2013eca,Lindner:2010wr,Hashimoto:2011tn,Ding:2011cm,Morisi:2007ft,Adhikary:2008au,Lin:2009bw,Csaki:2008qq,Hagedorn:2009jy,Burrows:2009pi,Ding:2009gh,Mitra:2009jj,delAguila:2010vg,Burrows:2010wz,Ahn:2014zja,Karmakar:2014dva,Ahn:2014gva,He:2006dk,Berger:2009tt,Kadosh:2010rm,Lavoura:2012cv,King:2012in,Adulpravitchai:2009gi,Dorame:2011eb,Dorame:2012zv}, which we analyze in the following.
Their parameters are $C_1,~C_2\sim \mathcal{O}(1),~d=\pm1,\pm 1/2,$ and $ \chi_{1},\chi_2=0,\pi,\pm \pi/2$. We split our analysis into mass sum rules  for Majorana neutrinos and mass sum rules for Dirac neutrinos, followed by a discussion on how to differentiate between the two mass sum rules assuming either Majorana or Dirac neutrinos.
\subsection{Predictions for Majorana neutrinos}
\begin{figure}
    \includegraphics[width=\textwidth]{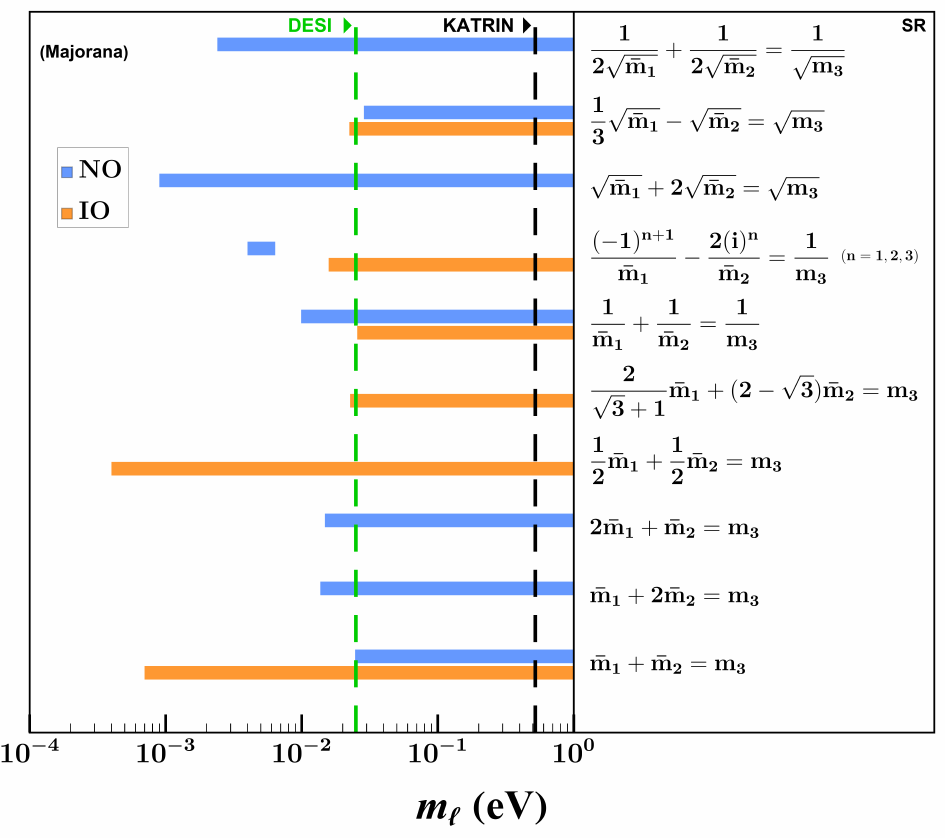}
    \caption{Predictions on $m_\ell$ for all valid mass sum rules for Majorana neutrinos found in the literature. The blue (orange) bands correspond to NO (IO).  The green line represents the $3\sigma$ constraint on the lightest neutrino mass in the NO from DESI \cite{DESI:2024hhd}, and the gray line represents masses disfavored by KATRIN at $3\sigma$ for either MO \cite{KATRIN:2024cdt}.
    }
    \label{fig:MajoranaSumRuleBandPlot}
\end{figure}

All 12 model predictions found in the literature can be realized for Majorana neutrinos however, only six model predictions are valid in both MOs, four are only valid in NO, and two only in IO.
Using eq.~\eqref{eq:GeneralSumRule} we obtain constraints on the lightest mass and the mass splittings. However, since the mass splittings are already precisely measured, see table \ref{tab:OscillationParameters}, we find in general that the uncertainty of the mass splittings plays a minimal role in the predictions of the lightest mass.  Furthermore, we don't find correlations or constrained  regions for the mass splittings, and hence we report our findings in terms of the predictions for the lightest mass only   in fig.~\ref{fig:MajoranaSumRuleBandPlot}. Three mass sum rules differ by just their phases $\chi_1,~\chi_2$ which don't play a role for the predictions on the lightest mass, see eqs.~\eqref{eq:TriangleInequalityNO}, \eqref{eq:TriangleInequalityIO}.

No mass sum rule predicts a lightest mass of zero which necessitates very precise values of $C_1,~C_2$, not found in the literature. In NO, this would require $C_2 = \left(\frac{\Delta m_{31}^2}{\Delta m_{21}^2}\right)^{d/2}$, and in IO $C_1/C_2 = \left(\frac{\left|\Delta m_{32}^2\right|}{\left|\Delta m^2_{32}\right|-\Delta m_{21}^2}\right)^{d/2}$. For all but three mass sum rules, there is no upper limit on the lightest mass since the Majorana phases can be adjusted as the lightest mass grows to complete the triangle. This means that all values of $m_\ell$ above the minimal value are equally preferred by the model prediction. 
The exceptions are the three mass sum rules with $\left\{C_1=1,C_2=2,d=-1\right\}$ for which in NO, $m_1^{-1}>m_2^{-1}>m_3^{-1}$, and so only a small range of masses is predicted where $m_3^{-1} \geq \left|m_1^{-1}-2 m_2^{-1}\right|$, leading to a narrow prediction range around $m_1 \sim \sqrt{\Delta m_{21}^2 /3}$. As discussed in sec.~\ref{sec:methodology} we 
limit the plot of our predictions to $m_\ell<1$ eV.

Generally, the mass sum rules we study favor large values of the lightest neutrino mass of around $m_\ell \gtrsim10^{-2}$ eV with few exceptions. Most model predictions contain a lower limit on the lightest mass which is allowed by the cosmological bound on the sum of neutrino masses from DESI \cite{DESI:2024hhd}, with the exception of $\tilde{m}_1+\tilde{m}_2 = m_3$ and $\frac{1}{3}\sqrt{\tilde{m}_1}-\sqrt{\tilde{m}_2} = \sqrt{m_3}$ in NO, whose lower limit on $m_\ell$ is just above  the cosmological bound.

\begin{figure}
    \includegraphics[width=\textwidth]{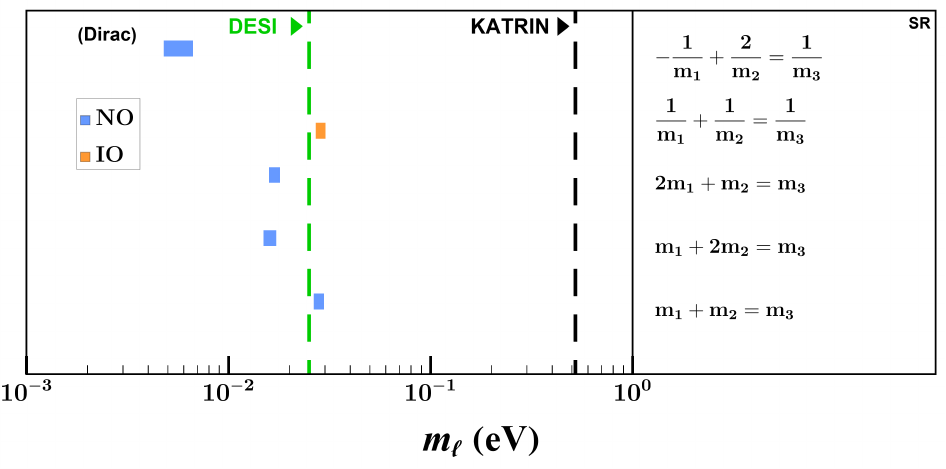}
    \caption{Predictions for the lightest mass assuming the mass sum rules for Dirac neutrinos. The mass sum rules not shown are  invalid. The color coding is the same as in fig.~\ref{fig:MajoranaSumRuleBandPlot}. 
    }
    \label{fig:DiracSumRuleBandPlot}
\end{figure}

\subsection{Predictions for Dirac neutrinos}
There is just one mass sum rule for Dirac neutrinos in the literature with values $C_1=C_2=1/2,~\chi_1=\pi=\chi_2,~d=1$ \cite{Ding:2013eca}. As we remain agnostic about the underlying model, we also consider mass sum rules for Dirac neutrinos which feature the same coefficients as for Majorana neutrinos. Out of the 12 mass sum rules studied, we find that for Dirac neutrinos, only five can be realized.  In particular, as we have defined the sum rule in eq.~\eqref{eq:GeneralSumRule} to involve the physical neutrino mass eigenvalues where unphysical phases are already absorbed, instead of the fully complex eigenvalues, we find that the Dirac mass sum rule in the literature cannot be fulfilled in our framework. This is in contrast to the original reference \cite{Ding:2013eca} where it was instead assumed that the mass sum rule involves the complex neutrino mass eigenvalues which introduces three complex phases and enough freedom to satisfy the mass sum rules.

In  fig.~\ref{fig:DiracSumRuleBandPlot} we show the predictions of the valid sum rules for the lightest neutrino masses. Again, we report no substantial correlations between $m_\ell$ and the mass splittings. As expected, the valid Dirac predictions lie within the corresponding predictions of the Majorana case.
Unlike the Majorana case, we find rather narrow allowed mass regions for Dirac neutrinos, with a lower and an upper limit on the lightest mass. All of the five valid mass sum rules predict just one MO, with only $1/m_1+1/m_2=1/m_3$ predicting the IO. 
Of the four valid in NO, three model predictions include a mass range in agreement with the DESI bound, with just $m_1+m_2 = m_3$ in contention.

\subsection{Differentiating models}
From figs.~\ref{fig:MajoranaSumRuleBandPlot} and \ref{fig:DiracSumRuleBandPlot}, we see that the predictions of mass sum rules in the current literature predict rather large masses around $m_\ell\gtrsim 10^{-2}$ eV with few exceptions. This means that these model predictions are testable with current and upcoming probes of the absolute neutrino mass scale.

\begin{table}
  \centering
  \caption{Overlap value $R$, defined in eq.~\eqref{eq:IntegralOverlap},  between valid mass sum rules for Majorana neutrinos assuming NO is true. 
  We classify the mass sum rules according to their parameters $\left\{C_1, C_2, d\right\}$ as defined by eq.~\eqref{eq:GeneralSumRule}. The exact ranges of masses from each model prediction are shown in fig.~\ref{fig:MajoranaSumRuleBandPlot}. The columns represent the ``true'' models and rows the ``test'' models. Model prediction pairs with an overlap of $0$ are completely disjoint in their predictions for $m_1$ and thus can always be differentiated while model prediction pairs with an overlap of $1$ can never be distinguished using measurements of $m_\ell$. }
 \begin{threeparttable}
    \definecolor{grey}{HTML}{9B9B9B}
    
    \renewcommand{\arraystretch}{1.3}
    \setlength{\tabcolsep}{5pt}

    \resizebox{\textwidth}{!}{
    \begin{tabular}{l ccccccccc}
    \toprule
    \toprule
    \rowcolor{grey}
    \diagbox{\small \textbf{TEST}}{\small \textbf{TRUE}} & 
    $\{1,1,1\}$ & 
    $\{1,2,1\}$ & 
    $\{2,1,1\}$ & 
    $\{1,1,-1\}$ & 
    $\{1,2,-1\}$ & 
    $\{1,2,\frac{1}{2}\}$ & 
    $\{\frac{1}{3},1,\frac{1}{2}\}$ &
    $\{\frac{1}{2},\frac{1}{2},-\frac{1}{2}\}$ \\
    
    \midrule

    $\{1,1,1\}$ 
    & --- 
    & \cellcolor{myorange}\textbf{0.80} 
    & \cellcolor{myorange}\textbf{0.81} 
    & \cellcolor{myorange}\textbf{0.70} 
    & \cellcolor{mygreen}\textbf{0} 
    & \cellcolor{myorange}\textbf{0.76}
    & \cellcolor{darkred}\textbf{0.97}
    & \cellcolor{myorange}\textbf{0.58} \\
    
    $\{1,2,1\}$ 
    & \cellcolor{darkred}{\textbf{1}} 
    & --- 
    & \cellcolor{darkred}{\textbf{1}} 
    & \cellcolor{darkred}{\textbf{0.88}} 
    & \cellcolor{mygreen}\textbf{0} 
    & \cellcolor{darkred}{\textbf{0.96}}
    & \cellcolor{darkred}\textbf{1}
    & \cellcolor{myorange}\textbf{0.72} \\
    
    $\{2,1,1\}$ 
    & \cellcolor{darkred}{\textbf{1}} 
    & \cellcolor{darkred}{\textbf{0.98}} 
    & --- 
    & \cellcolor{darkred}{\textbf{0.86}} 
    & \cellcolor{mygreen}\textbf{0} 
    & \cellcolor{darkred}{\textbf{0.94}}
    & \cellcolor{darkred}\textbf{1}
    & \cellcolor{myorange}\textbf{0.71} \\
    
    $\{1,1,-1\}$ 
    & \cellcolor{darkred}{\textbf{1}} 
    & \cellcolor{darkred}{\textbf{1}} 
    & \cellcolor{darkred}{\textbf{1}} 
    & --- 
    & \cellcolor{mygreen}\textbf{0} 
    & \cellcolor{darkred}{\textbf{1}}
    & \cellcolor{darkred}\textbf{1}
    & \cellcolor{myorange}\textbf{0.82} \\
    
    $\{1,2,-1\}$ 
    & \cellcolor{mygreen}\textbf{0} 
    & \cellcolor{mygreen}\textbf{0} 
    & \cellcolor{mygreen}\textbf{0} 
    & \cellcolor{mygreen}\textbf{0} 
    & --- 
    & \cellcolor{mygreen}\textbf{0}
    & \cellcolor{mygreen}\textbf{0}
    & \cellcolor{mygreen}\textbf{0.02} \\

    $\{1,2,\frac{1}{2}\}$ 
    & \cellcolor{darkred}{\textbf{1}} 
    & \cellcolor{darkred}{\textbf{1}} 
    & \cellcolor{darkred}{\textbf{1}} 
    & \cellcolor{darkred}{\textbf{0.92}} 
    & \cellcolor{mygreen}\textbf{0} 
    & --- 
    & \cellcolor{darkred}\textbf{1}
    & \cellcolor{myorange}\textbf{0.76} \\

    $\{\frac{1}{3},1,\frac{1}{2}\}$ 
    & \cellcolor{darkred}\textbf{0.98}
    & \cellcolor{myorange}\textbf{0.78}
    & \cellcolor{myorange}\textbf{0.80}
    & \cellcolor{myorange}\textbf{0.70}
    & \cellcolor{mygreen}\textbf{0}
    & \cellcolor{myorange}\textbf{0.74}
    & ---
    & \cellcolor{myorange}\textbf{0.55} \\
    
    $\{\frac{1}{2},\frac{1}{2},-\frac{1}{2}\}$ 
    & \cellcolor{darkred}{\textbf{1}} 
    & \cellcolor{darkred}{\textbf{1}} 
    & \cellcolor{darkred}{\textbf{1}} 
    & \cellcolor{darkred}{\textbf{1}} 
    & \cellcolor{mygreen}\textbf{0.12} 
    & \cellcolor{darkred}{\textbf{1}}
    & \cellcolor{darkred}\textbf{1}
    & --- \\
    
    \bottomrule
    \bottomrule
    \end{tabular}
    }
\end{threeparttable}
  \label{tab:SumRuleCompatibilityNO}
\end{table}

\begin{table}
\centering
    \caption{Overlap value $R$, defined in eq.~\eqref{eq:IntegralOverlap},  between valid mass sum rules for Majorana neutrinos assuming IO is true. The ``true'' models are represented in the columns, the ``test'' models in the rows,  
  according to their parameters $\left\{C_1, C_2, d\right\}$ defined in eq.~\eqref{eq:GeneralSumRule}. 
  See table~\ref{tab:SumRuleCompatibilityNO} for the corresponding results in NO.}
\begin{threeparttable}
    \definecolor{grey}{HTML}{9B9B9B} 
    
    \renewcommand{\arraystretch}{1.2}
    \setlength{\tabcolsep}{5pt}

\resizebox{\textwidth}{!}{
    \begin{tabular}{l cccccc}
    \toprule
    \toprule
    \rowcolor{grey}
    \diagbox{\small \textbf{TEST}}{\small \textbf{TRUE}} & 
    $\{1,1,1\}$ & 
    $\{\frac{1}{2},\frac{1}{2},1\}$ & 
    $\{\frac{2}{\sqrt{3}+1}, 2-\sqrt{3}, 1\}$ & 
    $\{1,1,-1\}$ & 
    $\{1,2,-1\}$ &
    $\{\frac{1}{3},1,\frac{1}{2}\}$ \\
    
    \midrule

    $\{1,1,1\}$ 
    & --- 
    & \cellcolor{darkred}\textbf{0.90} 
    & \cellcolor{darkred}\textbf{1} 
    & \cellcolor{darkred}\textbf{1} 
    & \cellcolor{darkred}\textbf{1}
    & \cellcolor{darkred}\textbf{1} \\
    
    $\{\frac{1}{2},\frac{1}{2},1\}$ 
    & \cellcolor{darkred}\textbf{1} 
    & --- 
    & \cellcolor{darkred}\textbf{1} 
    & \cellcolor{darkred}\textbf{1} 
    & \cellcolor{darkred}\textbf{1}
    & \cellcolor{darkred}\textbf{1} \\
    
    $\{\frac{2}{\sqrt{3}+1}, 2-\sqrt{3}, 1\}$ 
    & \cellcolor{myyellow}\textbf{0.39} 
    & \cellcolor{myyellow}\textbf{0.35} 
    & --- 
    & \cellcolor{darkred}\textbf{1} 
    & \cellcolor{darkred}\textbf{0.86}
    & \cellcolor{darkred}\textbf{1} \\
    
    $\{1,1,-1\}$ 
    & \cellcolor{myyellow}\textbf{0.37} 
    & \cellcolor{myyellow}\textbf{0.33} 
    & \cellcolor{darkred}\textbf{0.95} 
    & --- 
    & \cellcolor{darkred}\textbf{0.82}
    & \cellcolor{darkred}\textbf{0.95} \\
    
    $\{1,2,-1\}$ 
    & \cellcolor{myyellow}\textbf{0.45} 
    & \cellcolor{myyellow}\textbf{0.40} 
    & \cellcolor{darkred}\textbf{1} 
    & \cellcolor{darkred}\textbf{1} 
    & ---
    & \cellcolor{darkred}\textbf{1} \\

    $\{\frac{1}{3},1,\frac{1}{2}\}$ 
    & \cellcolor{myyellow}\textbf{0.39}
    & \cellcolor{myyellow}\textbf{0.34}
    & \cellcolor{darkred}\textbf{1}
    & \cellcolor{darkred}\textbf{1}
    & \cellcolor{darkred}\textbf{0.85}
    & --- \\
    
    \bottomrule
    \bottomrule
    \end{tabular}
}    
\end{threeparttable}

\label{tab:SumRuleCompatibilityIO}
\end{table}

As many of the model predictions include similar values for the lightest mass and show no strong preferences for any value of the mass splittings within their experimentally preferred ranges, it is challenging to differentiate the majority of mass sum rules from each other. We quantify the 
overlap as defined in eq.~\eqref{eq:IntegralOverlap} between each pairing of mass sum rules 
for Majorana neutrinos in each MO in tables \ref{tab:SumRuleCompatibilityNO} and \ref{tab:SumRuleCompatibilityIO}. In NO, we find that only the mass sum rule with $(-1)^{n+1}/\tilde m_1+2(\text{i})^n/\tilde m_2=1/\tilde m_3~,(n=1,2,3)$  can be easily distinguished from other mass sum rules due to its upper bound on $m_1$, while the all other pairs of mass sum rules feature an overlap of 50\% or more. In IO, no mass sum rule can be easily distinguished from all the others, but several model prediction pairs feature only a moderate overlap of $\sim 30\%-50\%$ with the majority of model prediction pairs exhibiting a large overlap $\gtrsim 80\%$.

We do not report the overlap score for Dirac neutrinos since the mass sum rules make narrow, non-overlapping predictions on the lightest mass, with the exception of mass sum rules $m_1+2m_2=m_3$ and $2m_1+m_2=m_3$ in NO.

\subsection{Discussion}
To summarize, neutrino mass sum rule model predictions arise in models where the three neutrino masses are parametrized with two parameters only. The resulting relation between the neutrino masses is of the form in eq.~\eqref{eq:GeneralSumRule} with $C_1,~C_2,~d$ constants predicted by the underlying model. The only measurable prediction of neutrino mass sum rules is the lightest neutrino mass (not counting the Majorana phases which can only be partially constrained), where we find that existing neutrino mass sum rules generally predict large lightest masses of $m_\ell\gtrsim 10^{-2}$ eV, making them testable in the near future by beta decay endpoint spectrum measurements and cosmological constraints on the sum of neutrino masses. However, these measurements do not allow us to distinguish
most neutrino mass sum rules for Majorana neutrinos from each other, as only a lower limit on the lightest mass is predicted. For Dirac neutrinos, mass sum rules however lead to both an upper and lower limit on the lightest mass and relatively narrow mass predictions, allowing generally for distinctions between different model predictions based on neutrino mass measurements only.

\section{Texture-zeros}
\label{sec:texture zeros}
This class of flavor models features neutrino mass matrices where one or two elements are equal to zero. We define the neutrino mass matrix elements as 
\begin{equation}
    M_\nu \equiv \begin{pmatrix}
m_{ee} & m_{e\mu} & m_{e\tau}\\
m_{\mu e} & m_{\mu\mu} & m_{\mu\tau}\\
m_{\tau e}&m_{\tau \mu}&m_{\tau\tau}
\end{pmatrix}~.
\label{eq:MajMassMatrixDefinition}
\end{equation}
Depending on the nature of neutrinos, the mass matrix is diagonalized by $U^\dagger M_\nu U=M_\nu^\text{diag}$ (Dirac) or  $U^T M_\nu U=M_\nu^\text{diag}$ (Majorana)
with the PMNS matrix $U$ \cite{Pontecorvo:1957cp,Maki:1962mu} introduced in sec.~\ref{sec:model_overview}
and $M_\nu^\text{diag}$ contains the neutrino mass eigenvalues, including the Majorana phases in the Majorana neutrino case. The equation for the zero matrix element $m_{ij}$ then leads 
to a sum-rule-style expression as in eq.~\eqref{eq:GeneralSumRule}. 
However, in this case the coefficients $C_i$ are now functions of the PMNS matrix elements and, as this is a linear equation, $d=1$. 
Hence, the predictions on the lightest mass and the mass splittings now depend on the mixing parameters.
This section is focused on finding and analyzing the correlations among them.
For Majorana neutrinos, the matrix element $m_{ee}=0$ is of particular interest as it corresponds to the observable in 0$\nu\beta\beta$. In particular, the case of $m_{ee}=0$ corresponds to a vanishing 0$\nu\beta\beta$ rate, even when neutrinos are Majorana.

Texture-zero matrices can arise in models based on an extended scalar sector and suitable Abelian symmetries \cite{Grimus:2004hf}.
Here we study Majorana and Dirac neutrinos for the cases of one texture-zero and two texture-zeros only, as more zero matrix elements do not have enough freedom to reproduce all mixing parameters \cite{Frampton:2002yf}.
We relate the common notation for two texture-zeros of \cite{Fritzsch:2011qv} to our notation in table \ref{tab:texture two zeros}.

\begin{table}
\centering
\caption{The dictionary between the oft-used notation of \cite{Fritzsch:2011qv} and our notation for two texture-zero model predictions.
The top two are the only two viable with the cosmology constraint for either Majorana or Dirac.}
\begin{tabular}{c|c}
$A_1$ & $m_{ee}=m_{e\mu}=0$\\
$A_2$ & $m_{ee}=m_{e\tau}=0$\\\hline
$B_1$ & $m_{\mu\mu}=m_{e\tau}=0$\\
$B_2$ & $m_{\tau\tau}=m_{e\mu}=0$\\
$B_3$ & $m_{\mu\mu}=m_{e\mu}=0$\\
$B_4$ & $m_{\tau\tau}=m_{e\tau}=0$\\
$C$ & $m_{\mu\mu}=m_{\tau\tau}=0$\\
$D_1$ & $m_{\mu\mu}=m_{\mu\tau}=0$\\
$D_2$ & $m_{\tau\tau}=m_{\mu\tau}=0$\\
$E_1$ & $m_{ee}=m_{\mu\mu}=0$\\
$E_2$ & $m_{ee}=m_{\tau\tau}=0$\\
$E_3$ & $m_{ee}=m_{\mu\tau}=0$\\
$F_1$ & $m_{e\mu}=m_{e\tau}=0$\\
$F_2$ & $m_{e\mu}=m_{\mu\tau}=0$\\
$F_3$ & $m_{e\tau}=m_{\mu\tau}=0$\\
\end{tabular}
\label{tab:texture two zeros}
\end{table}
\subsection{Majorana neutrinos}
For Majorana neutrinos, we define the mass matrix in the flavor basis as
\begin{equation}
    M_\nu =  U \diag(m_1 e^{i\alpha}, m_2 e^{i\beta},m_3) U^T,
\label{eq:MajMatrixDefinition}
\end{equation}
where $\alpha,\beta$ are again the Majorana phases, $m_1,m_2,m_3\geq0$ the neutrino mass eigenvalues. Note that the Majorana mass matrix is a complex, symmetric matrix and hence $m_{ij} = m_{ji}.$

\subsubsection{One texture-zeros}
Setting one element of eq.~\eqref{eq:MajMatrixDefinition} to $m_{i j} = 0$ where $i,j\in \{e,\mu,\tau\}$ yields the expression
\begin{equation}
    U_{i1}U_{j1} m_1 e^{i\alpha}+U_{i2}U_{j2}m_2 e^{i\beta}+U_{i3}U_{j3}m_3 = 0\,,
    \label{eq:MajoranaTZSumRule}
\end{equation}
which is of the same form as eq.~\eqref{eq:GeneralSumRule} when we identify 
\begin{equation}
    C_{1} = \left|\frac{U_{i1}U_{j1}}{U_{i3}U_{j3}}\right|,\;\;\; C_2 = \left|\frac{U_{i2}U_{j2}}{U_{i3}U_{j3}}\right|,
    \label{eq:Majorana1TZCoefficients}
\end{equation}
and $d=1$.
We can absorb the phases $\chi_1$ and $\chi_2$ into the complex phases in the PMNS matrix elements. 
In particular, we observe a $\mu-\tau $ symmetry  at heavier masses as the precise values of the coefficients, which are functions of the mixing angles in this case, become less important. Additionally, the coefficients of the pairs of textures  $m_{e\mu}=0, m_{e\tau}=0$ and $m_{\mu\mu}=0$, $m_{\tau\tau}=0$ are related by an exchange of $c_{23}\to -s_{23}$,  $s_{23}\to c_{23}$  and $\delta\to \delta+\pi$ leading to their predictions on $\cos\delta$ being mirrored at $\cos\delta=0$ given the current lack of information on the octant.

We obtain six unique model predictions of one texture-zeros.
Each of them predicts correlations among the seven low-energy neutrino parameters with the exception of $m_{ee}=0$ which has no dependence on $\theta_{23}$ or $\delta$ in the usual parameterization.\footnote{In different parameterizations of the PMNS matrix, different one texture-zeros will be independent of some of the parameters of the matrix.
Thus the fact that $m_{ee}=0$ is special in depending on fewer of the parameters is not a fundamental statement about $m_{ee}$, although the fact that it predicts a vanishing 0$\nu\beta\beta$ rate does make $m_{ee}$ unique, as that is a physical observable.} Following the methodology explained in sec.~\ref{sec:methodology}, we find that all possible model predictions can account for the measured parameters \cite{Lashin:2011dn} however not all of them are valid in both MOs, see also \cite{Borah:2025vtn} for a recent study in light of JUNO data. Additionally, the model predictions include correlations between $m_\ell$, $\theta_{23}$, and $\cos\delta$, while the other neutrino parameters are not or very weakly correlated. The precision on the mass splittings plays a subdominant role in the predictions of other parameters.  

Noteworthy predictions in the  NO are: 
\begin{itemize}
\item{ While there are minor correlations between the lower bounds of $m_1$ and $\delta$, we note that for all one texture-zeros, these correlations only vary within a mass range of $\sim 2$ meV and are not prominent enough to  be  detectable by absolute mass scale experiments.}
\item{None of the model predictions include an upper bound on $m_{1}$, with the exception of $m_{ee}=0$, which predicts $m_1\in[2, 8]$ meV. This region corresponds to the vanishing 0$\nu\beta\beta$ rate. }
     \item{
   The $m_{ee}=0$ texture prefers $\theta_{12}$ at the upper end of its $3\sigma$ range, $\theta_{12}>33.5^\circ$ when $m_1\in[6,8]$ meV. 
     While this occurs in a narrow mass range,  this aligns with the range in which the mass predictions for $m_1$ of $m_{ee}=0$, and $m_{e\mu}=0$ or $m_{e\tau}=0$ intersect.}
    \item{The model predictions for $m_{\mu\mu}=0,~ m_{\mu\tau}=0$, and $m_{\tau\tau}=0$ for the minimum lightest mass are similar; all predict roughly $m_1 \gtrsim 30$ meV. However, they also all exhibit a preference for the octant of $\theta_{23}$: where $m_{\mu\mu}=0$ indicates a clear preference for the lower octant, and $m_{\mu\tau}=0$ and $m_{\tau\tau}=0$ show a clear preference for the upper octant. At higher mass scales over $m_1\gtrsim 100$ meV, all three are consistent with $\theta_{23}$ being maximal.} 
    \item{The model predictions of $m_{e\mu}=0$ and $m_{e\tau}=0$ on the minimum lightest mass are similar, with both predicting $m_1 \gtrsim 3$ meV. 
    However, in the range of $m_1 \in [3,8]$ meV, $m_{e\tau}=0$ displays a preference for the upper octant of $\theta_{23}$, while $m_{e\mu}=0$ prefers the lower octant. }
\end{itemize}
\begin{figure}
\centering
    \includegraphics[width = .95\textwidth]{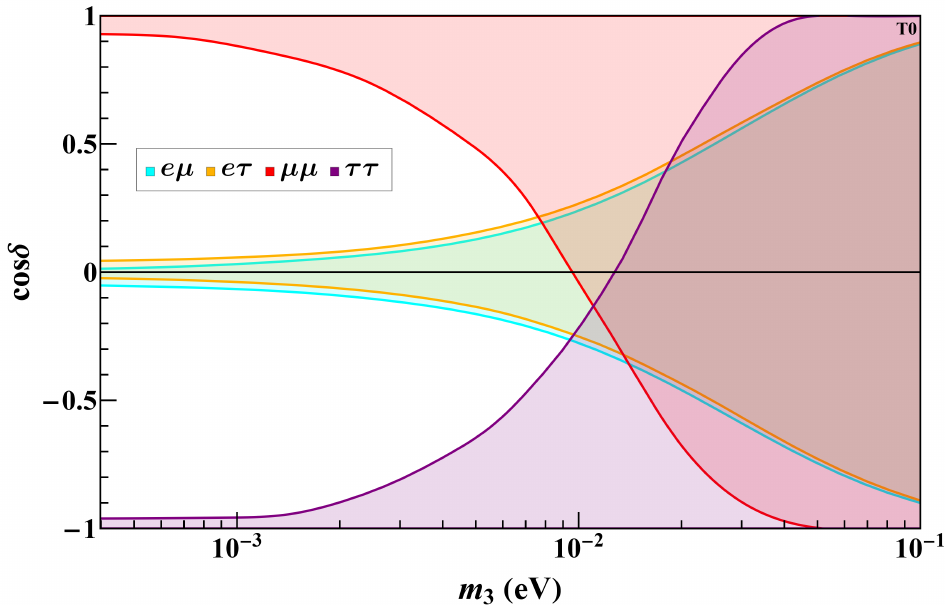}
    \caption{Predicted regions (shaded) in the $m_3$-$\cos\delta$ plane for the texture-zeros $m_{e\mu},~ m_{e\tau}, ~m_{\mu\mu},~m_{\tau\tau}=0$ in the IO assuming Majorana neutrinos. The contours correspond to $\Delta\chi^2=11.83$. We do not show $m_{ee}=0$ as is not allowed in IO. We also omit the preferred regions of $m_{\mu\tau}=0$  as the correlations between $m_3-\cos\delta$ are less pronounced. 
    As the mass scale of $m_3$ is pushed lighter, the permitted region of allowed $\delta$ decreases, making the model predictions more discernible at lower mass scales, with the exception of the pair $m_{e\mu}=0$ and $m_{e\tau}=0$ which make very similar predictions. In the NO, these correlations are much less pronounced and therefore we do not show them. }
    \label{fig:IOm3DeltaContours}
\end{figure}
The texture  $m_{ee}=0$ is not allowed in the IO. The other predictions in the IO are:
\begin{figure}
    \includegraphics[width = .95\textwidth]{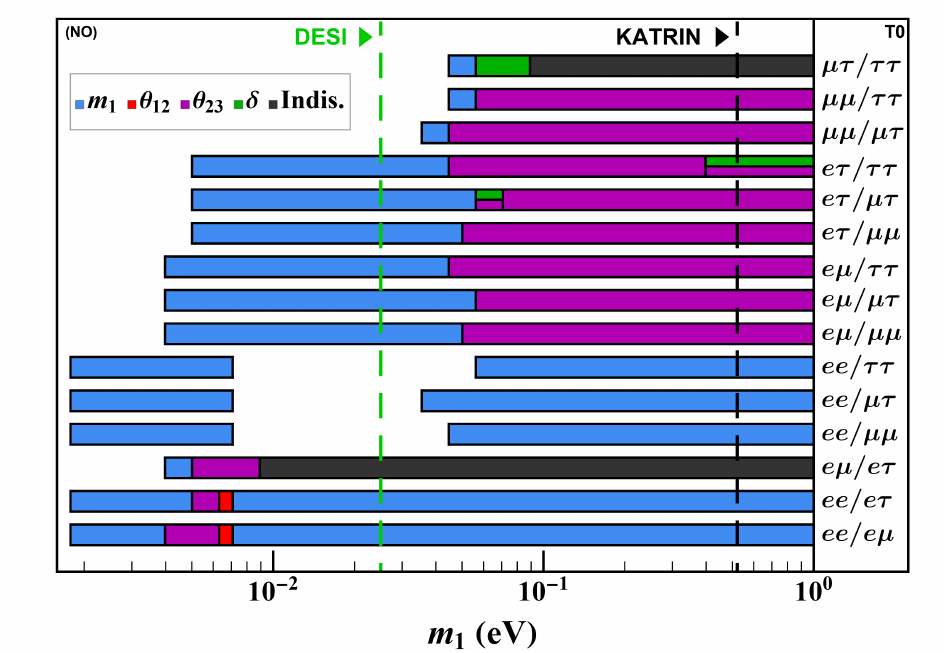}
    \caption{Identification of the most likely neutrino  observables to distinguish between two different one texture-zero model predictions in the NO assuming Majorana neutrinos. The model predictions are labeled according to the flavor indices of the zero mass matrix element. 
    Indication of $\theta_{23}$ (purple) is dependent on the resolution of the octant of $\theta_{23}$. The model predictions are indistinguishable (black) at certain mass scales if they cannot be differentiated using any mixing parameter measurements at that mass scale. Regions with two colors can use either parameter individually to distinguish the two model predictions. Not-shaded regions mean that both model predictions are not viable. None of the model predictions can be made consistent with $m_1=0$. We also show the $3\sigma$ constraints from DESI and KATRIN on the neutrino mass scale.} 
    \label{fig:NOMajDifferentiatorBands}
\end{figure}
\begin{itemize}
 \item{Model predictions in the IO exhibit much more notable correlations between $m_3$ and $\cos\delta$, as demonstrated in fig.~\ref{fig:IOm3DeltaContours} which can be tested by DUNE and HK in the future, combined with future measurements on the absolute neutrino mass scale.}
    \item{In general, the IO permits  much lower values for the lightest mass than in the NO. None of the model predictions include an upper bound on $m_3$, and again $m_{\mu\mu}=0$, $m_{\mu\tau}=0$, and $m_{\tau\tau}=0$ make similar predictions on a lower bound on the order of about $m_3 \gtrsim 0.5$ meV. }
    \item {The model predictions of $m_{e\tau}=0$ and $m_{e\mu}=0$ both permit $m_3=0$ within 3$\sigma$ values of the oscillation parameters. In the case of $m_3=0$, $\cos\delta$ is predicted to be within
\begin{align}
-0.050\leq& \cos\delta \leq 0.012 \qquad (m_{e\mu}=0)~,\\
-0.021\leq& \cos\delta \leq 0.040 \qquad (m_{e\tau}=0)~.
\end{align}
} 
\item{The predictions of $m_{e\mu}=0$ and $m_{e\tau}=0$ are very similar due to the approximate $\mu$-$\tau$ symmetry and their  predictions for
   $\cos\delta$ are mirrored at $\cos\delta=0$. However, as their predictions are narrow ranges around $\cos\delta\approx 0$, these model predictions are effectively indistinguishable at the current upcoming measurement prospects. Their predictions of $\delta$ at lower mass scales cannot be probed by the current sensitivities of DUNE and HK, and their predictions on $\theta_{23}$ are too close to maximal at low mass scales for them to be differentiated. Both will either be ruled out together or will both remain valid at the same confidence level.
    }
\end{itemize}

\begin{figure}
    \centering
    \includegraphics[width=\textwidth]{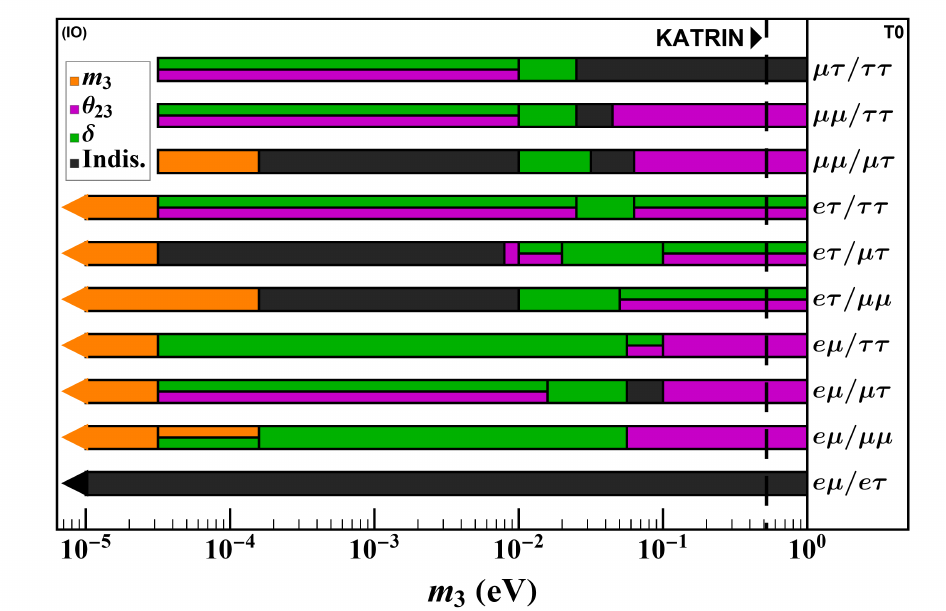}
    \caption{Identification of the most likely neutrino  observables to distinguish between two different one texture-zero model predictions in the IO assuming Majorana neutrinos. The model predictions are labeled according to the flavor indices of the zero mass matrix element. Bands with arrows continue on to $m_3 = 0$, see fig.~\ref{fig:NOMajDifferentiatorBands} for further explanations.}
    \label{fig:IOMajDifferentiatorBands}
\end{figure}

\subsubsection{Differentiating models}
In figs.~\ref{fig:NOMajDifferentiatorBands}, \ref{fig:IOMajDifferentiatorBands} we show the observable, given the expected sensitivity of next generation experiments,  is most likely  to differentiate between two model predictions.
We focus on one MO at a time, as the MO will be measured in a straightforward fashion with future experiments.
We then consider a measurement of the absolute mass scale from cosmology or direct searches, an octant determination (which may not be possible if $\theta_{23}$ is too close to $45^\circ$), a measurement of $\delta$, or a measurement of $\theta_{12}$.

We note several interesting features.
First, the observables have the most pronounced non-trivial dependence on the absolute neutrino mass scale which we hence chose as the x-axis.  
Second, in the NO the absolute mass scale and $\theta_{23}$ are most often the critical discriminators, while in the IO  $\cos\delta$  is often the strongest discriminator.
Third, we see that in the IO it is often not possible to discriminate between model predictions, but most can be discriminated in the NO.
Those that are hard to differentiate are often due to challenges in breaking approximate $\mu-\tau$ symmetries.

\begin{table}
\centering
\caption{Overlap $R$, as defined in eq.~\eqref{eq:IntegralOverlap}, for different one texture-zeros in NO assuming Majorana neutrinos, see table~\ref{tab:IOTexZeroMCIntegration} for the results in IO.
The format is the same as table~\ref{tab:SumRuleCompatibilityNO}.
}
\begin{threeparttable}
    \definecolor{grey}{HTML}{9B9B9B}
    
    \renewcommand{\arraystretch}{1.3}
    \setlength{\tabcolsep}{10pt}
    
    \begin{tabular}{l cccccc}
    \toprule
    \toprule
    \rowcolor{grey}
    \diagbox{\small \textbf{TEST}}{\small \textbf{TRUE}} & 
    $m_{ee}$ & $m_{e\mu}$ & $m_{e\tau}$ & $m_{\mu\mu}$ & $m_{\mu\tau}$ & $m_{\tau\tau}$ \\
    \midrule

    $m_{ee}$      & --- & \cellcolor{mygreen}\textbf{0.08} & \cellcolor{mygreen}\textbf{0.24} & \cellcolor{mygreen}\textbf{0} & \cellcolor{mygreen}\textbf{0} & \cellcolor{mygreen}\textbf{0} \\
    
    $m_{e\mu}$     & \cellcolor{mygreen}\textbf{0.03} & --- & \cellcolor{darkred}\textbf{0.97} & \cellcolor{mygreen}\textbf{0.08} & \cellcolor{mygreen}\textbf{0.17} & \cellcolor{mygreen}\textbf{0.23} \\
    
    $m_{e\tau}$    & \cellcolor{mygreen}\textbf{0.09} & \cellcolor{darkred}\textbf{0.89} & --- & \cellcolor{mygreen}\textbf{0.07} & \cellcolor{mygreen}\textbf{0.02} & \cellcolor{mygreen}\textbf{0.21} \\
    
    $m_{\mu\mu}$   & \cellcolor{mygreen}\textbf{0} & \cellcolor{darkred}\textbf{1} & \cellcolor{darkred}\textbf{1} & --- & \cellcolor{mygreen}\textbf{$<$ 0.01} & \cellcolor{mygreen}\textbf{$<$ 0.01} \\
    
    $m_{\mu\tau}$  & \cellcolor{mygreen}\textbf{0} & \cellcolor{darkred}\textbf{1} & \cellcolor{darkred}\textbf{1} & \cellcolor{mygreen}\textbf{$<$ 0.01} & --- & \cellcolor{darkred}\textbf{1} \\
    
    $m_{\tau\tau}$ & \cellcolor{mygreen}\textbf{0} & \cellcolor{darkred}\textbf{1} & \cellcolor{darkred}\textbf{1} & \cellcolor{mygreen}\textbf{$<$ 0.01} & \cellcolor{myorange}\textbf{0.70} & --- \\
    
    \bottomrule
    \bottomrule
    \end{tabular}
\end{threeparttable}
\label{tab:NOTexZeroMCIntegration}
\end{table}

\begin{table}
\caption{Overlap $R$, as defined in eq.~\eqref{eq:IntegralOverlap}, for different one texture-zeros in IO for Majorana neutrinos, see table~\ref{tab:NOTexZeroMCIntegration} for the results in NO.
The format is the same as table~\ref{tab:SumRuleCompatibilityNO}.
}
\centering
\begin{threeparttable}
    \definecolor{grey}{HTML}{9B9B9B}
    \definecolor{myyellow}{HTML}{FFFF00}
    \definecolor{darkred}{rgb}{0.85,0,0} 
    \definecolor{myorange}{HTML}{FFA500}
    
    \renewcommand{\arraystretch}{1.3}
    \setlength{\tabcolsep}{10pt}
    
    \begin{tabular}{l ccccc}
    \toprule
    \toprule
    \rowcolor{grey}
    \diagbox{\small \textbf{TEST}}{\small \textbf{TRUE}} & 
    $m_{e\mu}$ & $m_{e\tau}$ & $m_{\mu\mu}$ & $m_{\mu\tau}$ & $m_{\tau\tau}$ \\
    \midrule

    $m_{e\mu}$     & --- & \cellcolor{darkred}\textbf{0.97} & \cellcolor{myyellow}\textbf{0.38} & \cellcolor{darkred}\textbf{0.77} & \cellcolor{myorange}\textbf{0.71} \\
    
    $m_{e\tau}$    & \cellcolor{darkred}\textbf{0.97} & --- & \cellcolor{myyellow}\textbf{0.40} & \cellcolor{darkred}\textbf{0.79} & \cellcolor{myorange}\textbf{0.72} \\
    
    $m_{\mu\mu}$   & \cellcolor{myyellow}\textbf{0.31} & \cellcolor{myyellow}\textbf{0.31} & --- & \cellcolor{myorange}\textbf{0.61} & \cellcolor{myyellow}\textbf{0.31} \\
    
    $m_{\mu\tau}$  & \cellcolor{myorange}\textbf{0.56} & \cellcolor{myorange}\textbf{0.57} & \cellcolor{myorange}\textbf{0.51} & --- & \cellcolor{myorange}\textbf{0.64} \\
    
    $m_{\tau\tau}$ & \cellcolor{myorange}\textbf{0.58} & \cellcolor{myorange}\textbf{0.58} & \cellcolor{myyellow}\textbf{0.32} & \cellcolor{myorange}\textbf{0.73} & --- \\
    
    \bottomrule
    \bottomrule
    \end{tabular}
\end{threeparttable}
\label{tab:IOTexZeroMCIntegration}
\end{table}

In tables \ref{tab:NOTexZeroMCIntegration} and \ref{tab:IOTexZeroMCIntegration}, we show the overlap between two different one texture-zero model predictions using the measure introduced in eq.~\eqref{eq:IntegralOverlap} for both MOs. 
Overall, 20 out of 30  of the model prediction comparisons can be distinguished at better than 50\% in NO while there is generally more overlap between model prediction comparisons in IO: only six out of 20 of the pairs can be distinguished at more than 50\%.
By design, some of the pairs are slightly asymmetric, depending on which model prediction is taken to be ``true''. Note that there are fewer model prediction pairs in the IO since $m_{ee}=0$ is not viable in the IO.

\subsubsection{Two texture-zero predictions }
There are 15 two texture-zeros which impose two simultaneous conditions of the form of eq.~\eqref{eq:MajoranaTZSumRule}. This further restricts the number of free parameters. Furthermore, the only valid two texture-zeros are those where both one texture-zero conditions are valid.

Out of the 15 possible two texture-zeros, we find that only two are consistent with all the data including cosmology \cite{Denton:2023hkx,Alcaide:2018vni}.
If one relaxes the cosmology bound, then there are seven valid two texture-zeros in total, two of which are only valid in the NO and the other five valid in both MOs \cite{Frampton:2002yf,Guo:2002ei,Xing:2002ta,Desai:2002sz,Xing:2002ap,Dev:2006if,Dev:2006xu, Fritzsch:2011qv,Honda:2003pg}.
For the five model predictions permitted in both MOs, all include a lower bound on the lightest mass to be at least greater than $25$ meV, which causes them to be in tension with cosmological data.
The remaining two model predictions valid only in NO include a lower mass bound of $m_1 \gtrsim 3$ meV and are bounded from above $m_1\lesssim8$ meV: these are $m_{ee} = m_{e\mu}=0$, and $m_{ee}=m_{e\tau}=0$. The values of $m_1$ predicted for each of these two texture-zeros can be expressed analytically in terms of the other mixing parameters \cite{Frampton:2002yf,Fritzsch:2011qv}.
These expressions are the only two with easily tractable expressions.
The case of $m_{ee}=m_{e\mu}=0$ requires a lightest mass of
\begin{equation}
    m_1^2 = 4\frac{\left( s^4_{13}c^2_{23}+ s^2_{13}s^2_{23}\right)\Delta m^2_{31}-c^2_{23} c^4_{13} s^2_{12} \Delta m^2_{21}}{3\cos(2\theta_{13})+ 2c^2_{13} \cos(2\theta_{23})-1}~,
    \label{eq:m_eem_emuMassPrediction}
\end{equation}
and  for $m_{ee}=m_{e\tau}=0$
\begin{equation}
  m_1^2 = 4\frac{\left( s^4_{13}s^2_{23}+ s^2_{13}c^2_{23}\right)\Delta m^2_{31}-s^2_{23} c^4_{13} s^2_{12} \Delta m^2_{21}}{3\cos(2\theta_{13})- 2c^2_{13} \cos(2\theta_{23})-1}~,
    \label{eq:m_eem_etauMassPrediction}
\end{equation}
\begin{figure}
    \centering
    \includegraphics[width=0.8\linewidth]{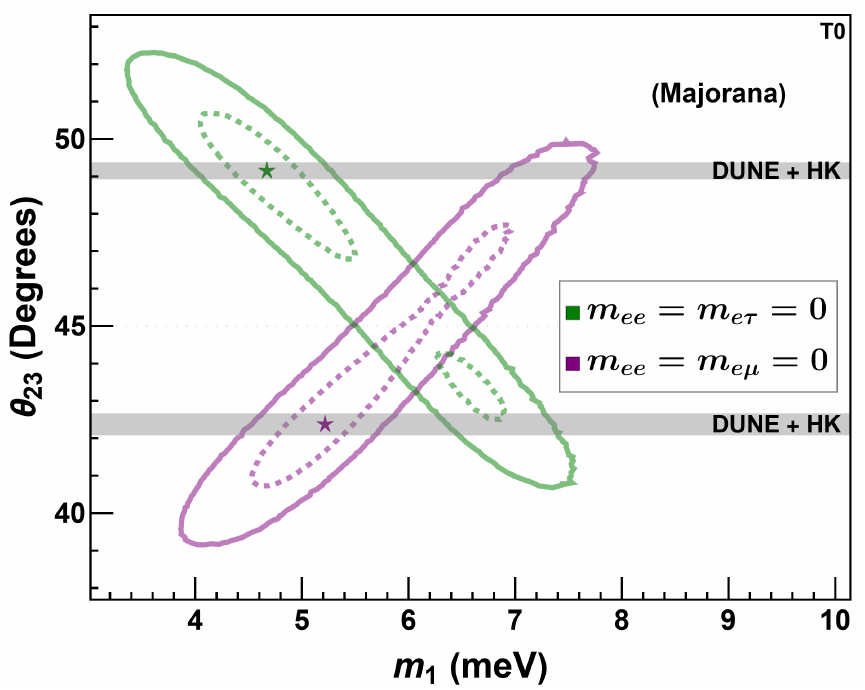}
    \caption{Predictions on $m_1,~\theta_{23}$ for the $m_{ee}=m_{e\mu}=0$ and  $m_{ee}=m_{e\tau}=0$ two texture-zeros in the NO for Majorana neutrinos. The  dotted and solid lines correspond to  $\Delta\chi^2 = 2.30$ and 11.83, we show the best-fit points with a star. The gray bands refer to the combined $1\sigma$ measurement resolution of DUNE and HK to $\theta_{23}$ at each of the best fit points, assuming reactor constraints on $\sin^2\theta_{13}$ and octant resolution is achieved. IO is not valid for these two texture-zero predictions.}
\label{fig:TwoTexZerosContourMajorana}
\end{figure} The predictions of these models are similar due to the approximate $\mu$-$\tau$ symmetry and because they come from the superposition of the one texture-zero predictions. These predictions have no preference for any value of $\theta_{12}, ~\theta_{23}$ or the mass splittings within their experimental $3\sigma $ regions.
However, these two model predictions include correlations between $m_1$ and $\theta_{23}$, as demonstrated by the contours in fig.~\ref{fig:TwoTexZerosContourMajorana}, which suggests that a precise measurement of $\theta_{23}$ by DUNE and HK, can be used to differentiate the two model predictions, depending on the absolute mass scale. We show the expected sensitivity from DUNE and HK for $\theta_{23}$ at each of the model prediction spaces' best-fit points.  

  For the remaining two texture-zeros valid in both MOs, we report the most notable predictions and correlations. In the NO:

  \begin{itemize}
      \item{
      We find that the predictions of $m_{e\mu}= m_{\mu\mu}=0$ and $m_{e\tau}=m_{\tau\tau}=0$ include a very similar lower bound of $m_1 \gtrsim 80$ meV, with the similarity in their mass prediction again coming from the approximate $\mu$-$\tau$ symmetry. Similarly, the model predictions of $m_{e\mu}=m_{\tau\tau}=0$ and $m_{e\tau}=m_{\mu\mu}=0$ constrain $m_1 \gtrsim 30$ meV. The final valid two texture-zero, $m_{\mu\mu}=m_{\tau\tau}=0$ predicts a higher upper bound of $m_1\gtrsim140$ meV. These higher mass scale predictions place most model predictions in contention with cosmology, but also makes them more probable by neutrino mass scale measurements.}
      \item{The two texture-zeros tend to make much more restrictive predictions on $\cos\delta$: $m_{e\mu}=m_{\mu\mu}=0$ predicts $\cos\delta\in[0.004, 0.022]$, and  $m_{e\tau}=m_{\tau\tau}=0$ predicts  $\cos\delta\in[-0.025, -0.002]$. Additionally, $m_{e\tau}=m_{\mu\mu}=0$ predicts $\cos\delta\in[-0.07,0.005]$ and $m_{e\mu}=m_{\tau\tau} = 0$ predicts $\cos\delta\in[-0.01,0.05]$. These correspond to a very narrow range within $2^\circ$ of $\delta\sim 270^\circ$, making these predictions difficult to distinguish, but easily ruled out by a measurement of $\delta\neq 270^\circ$. Finally, $m_{\mu\mu}=m_{\tau\tau}=0$ does not contain any notable predictions on $\delta$, allowing $\cos\delta \in[-1,1]$.}
      \item{The model predictions of $m_{e\mu}=m_{\mu\mu}=0$ and $m_{e\tau} = m_{\tau\tau}=0$ are constrained strictly to the lower and upper octant of $\theta_{23}$, respectively. The observables $m_1$ and $\theta_{23}$ are highly correlated such that a lower mass corresponds to a further deviation from $\theta_{23}=45^\circ$, with the deviation increasing to more than $2^\circ$ when $m_1\lesssim 120$~meV. There is a similar behavior in the model predictions for $m_{e\tau}=m_{\mu\mu}=0$ and $m_{e\mu}=m_{\tau\tau}=0$, which predict $\theta_{23}$ to be in its lower and upper octant, respectively, and whose deviation from maximal in $\theta_{23}$ becomes greater than $2^\circ$ when $m_1 \lesssim 60$ meV. In contrast, $m_{\mu\mu}=m_{\tau\tau}=0$ predicts $\theta_{23}\sim 45^\circ$ with less than $0.2^\circ$ deviation at all mass scales. A definitive octant measurement will be very effective at ruling out many of these model predictions.}
  \end{itemize}
In the IO, the two texture-zeros tend to behave very similarly to the NO. We report the most notable predictions and correlations in the IO:

\begin{itemize}
    \item{
    We find that the model predictions of $m_{e\mu}= m_{\mu\mu}=0$ and $m_{e\tau}=m_{\tau\tau}=0$ include a similar lower bound of $m_3 \gtrsim 20$ meV. The model predictions of $m_{e\mu}=m_{\tau\tau}=0$ and $m_{e\tau}=m_{\mu\mu}=0$ include the lower bound $m_3\gtrsim 40$ meV. The final valid two texture-zero in IO, $m_{\mu\mu}=m_{\tau\tau}=0$ predicts a higher upper bound of $m_3\gtrsim 50$ meV.}
    \item{The restrictive behavior on $\cos\delta$ is similar to the NO case. The model predictions for $m_{e\mu}=m_{\mu\mu}=0$ include $\cos\delta\in[-0.015,0.004]$, and similarly $m_{e\tau}=m_{\tau\tau}=0$ a range $\cos\delta\in[-0.008, -0.01]$. The model predictions for $m_{e\tau}=m_{\mu\mu}=0$ include $\cos\delta\in[0.004,0.007]$ and those for $m_{e\mu}=m_{\tau\tau} = 0$ a range $\cos\delta\in[-0.08,0.04]$. Again, this narrow range within $2^\circ$ of $\delta\sim 270^\circ$ makes these predictions difficult to distinguish through a measurement of $\delta$ alone but easily ruled out. Finally, $m_{\mu\mu}=m_{\tau\tau}=0$ does not make any notable predictions on $\delta$, with $\cos\delta\in [-1,1]$.}
     \item{The model predictions for $m_{e\mu}=m_{\mu\mu}=0$ and $m_{e\tau} = m_{\tau\tau}=0$ are restricted strictly to the upper and lower octant of $\theta_{23}$, respectively (note the switch of octant preference from NO). Again, $m_3$ and $\theta_{23}$ are highly correlated such that a lower mass corresponds to a further deviation from $\theta_{23}=45^\circ$, with the deviation increasing to more than $2^\circ$ when $m_1\lesssim60$ meV. There is a similar behavior for $m_{e\tau}=m_{\mu\mu}=0$ and $m_{e\mu}=m_{\tau\tau}=0$, which predict $\theta_{23}$ to be in its upper and lower octant, respectively, and whose deviation from maximal in $\theta_{23}$ becomes greater than $2^\circ$ when $m_3 \lesssim 100$ meV. Finally, $m_{\mu\mu}=m_{\tau\tau}=0$ makes no substantial predictions on the octant of $\theta_{23}$, but there is some correlation between $\theta_{23}$ and $\cos\delta$, with $\cos\delta>0$ tending to prefer the lower octant and $\cos\delta<0$ the upper octant. }
\end{itemize}

\subsection{Dirac neutrinos}
For Dirac neutrinos, the neutrino mass matrix is diagonalized with
\begin{equation}
    M_\nu =  U \diag(\eta m_1, \zeta m_2,m_3) U^\dagger,
\label{eq:DirMatrixDefinition}
\end{equation}
where $\eta,\zeta = \pm 1$ are the signs of each eigenvalue relative to the third, since a Hermitian matrix must have real but not necessarily positive eigenvalues \cite{Liu:2012axa}. Note that since this matrix is Hermitian, setting $m_{ij}=0$ is still equivalent to setting $m_{ji}= m_{ij}^*=0$, and there are still six unique one texture-zeros to consider. Setting one element $m_{ij}=0$ leads to
\begin{equation}
    \eta U_{i1}U^*_{j1} m_1+\zeta U_{i2}U^*_{j2}m_2+U_{i3}U^*_{j3}m_3 = 0\,,
    \label{eq:DiracTZSumRule}
\end{equation}
for $i,j\in \{e,\mu,\tau\}$. We note again that this matches the form of eq.~\eqref{eq:GeneralSumRule} with $d=1$ and the coefficients and phases given by the elements of $U$, like in eq.~\eqref{eq:Majorana1TZCoefficients}. Again, the predictions only depend on the absolute values of the coefficients $C_1$ and $C_2$ and not their phases. The Dirac case can again be seen as a special case of the Majorana case, and hence all Dirac neutrino observable predictions must lie within the corresponding Majorana prediction parameter space.
With the absence of the Majorana phases, the Dirac one texture-zeros are inherently more predictive due to the lower degrees of freedom. 
We note that eq.~\eqref{eq:DiracTZSumRule} represents two constraint equations only if part of the equation is complex. The off-diagonal Dirac texture-zeros predict precisely $\delta = 0,\pi$, which makes eq.~\eqref{eq:DiracTZSumRule} fully real and hence corresponds to only one constraint.

\subsubsection{One texture-zeros}
The 6 possible Dirac one texture-zeros exhibit distinct behaviors in their predictions on $\delta$, independent of MO. Those with zeros in the off-diagonal matrix elements, those being $m_{e\mu},m_{e\tau}$, $m_{\mu\tau} = 0$, impose a constraint equation which is generally complex and hence represent two constraints on the mixing parameters. However, the requirement that the imaginary part of the expression is satisfied requires either $|m_2|=|m_1|$, which is obviously not permitted by current data, or $\sin\delta = 0$. Hence, these three can easily be ruled out in either MO by a measurement of $\delta \not = 0,\pi$. For the texture-zeros with a zero entry on the  diagonal mass matrix elements, those being $m_{ee},m_{\mu\mu}$, $m_{\tau\tau} = 0$, the constraint equation becomes a fully real expression, and hence represents only one constraint. Unlike the off-diagonal texture-zeros, these are not predictive in $\delta$, allowing $\cos\delta \in[-1,1]$ with no noticeable correlations with other mixing parameters. We also note that the Dirac texture-zeros do impose an upper bound for their predictions on $m_\ell$, which was not generally true in the Majorana case. 

We report the model predictions split up by MO. Noteworthy predictions in the NO are:

\begin{itemize}
    \item{ We find that $m_{\mu\tau}=0$ cannot be reasonably satisfied for the mixing parameters within their experimental $3\sigma$ ranges.}
    \item{As stated, the Dirac texture-zeros have upper bounds on the lightest mass. We find $m_{ee}=0$ predicts $m_1\in [5,8]$ meV, and $m_{e\mu}=0$ and $m_{e\tau}=0$ predict $m_1\in [3, 10]$ meV. The predictions of $m_{\tau\tau}=0$ and $m_{\mu\mu}=0$ contain higher upper bounds with $m_1\in[4,250]$ meV.}
    \item{We find $m_{\mu\mu}=0$ and $m_{\tau\tau}=0$ predict $\theta_{23}$ to be in its lower and upper octant, respectively. The deviation from $\theta_{23} = 45^\circ$ is highly correlated with a lower mass scale $m_1$. At scales $m_1\lesssim10$ meV, the deviation from maximal $\theta_{23}$ becomes greater than $2^\circ$, making these model predictions easily differentiated by an octant resolution measurement at this mass scale.}
    \item{As with the Majorana case, $m_{ee}=0$ does not make any predictions on $\theta_{23}$ or $\delta$. While the model predictions on $\theta_{12}$ are within its full experimental bounds, we do report a correlation between $m_1$ and $\theta_{12}$, with a lower mass $m_1$ preferring a lower value of $\theta_{12}$. A more precise measurement of $\theta_{12}$ will further tighten the bound on $m_1$, but differentiating these predictions from the other Dirac one texture-zeros will prove difficult.}
\end{itemize}
Noteworthy predictions for Dirac one texture-zeros in the IO are:

\begin{itemize}
    \item{ As in the Majorana case, $m_{ee}=0$ is  ruled out by the current data. Additionally, we also rule out $m_{e\mu}=0$ and $m_{e\tau}=0$ in the IO.}
    \item{Of the remaining one texture-zeros, $m_{\mu\tau}=0$ predicts a lightest mass range $m_3 \in [14,30]$ meV, and  $m_{\mu\mu}=0$ and $m_{\tau\tau}=0$ predict a higher range of $m_3\in[40,300]$ meV.}
    \item{As in the NO case, $m_{\mu\mu}=0$ and $m_{\tau\tau}=0$ predict the upper and lower octant for $\theta_{23}$, respectively (note the flip from the NO results). Again, we find strong correlations between $m_3$ and $\theta_{23}$. At scales $m_3\lesssim10$ meV, the deviation from maximal $\theta_{23}$ becomes greater than $2^\circ$ and again makes these model predictions differentiable from each other.}
\end{itemize}

\subsubsection{Two texture-zeros}
Only three of the two texture-zeros agree with the observed mixing parameters for Dirac neutrinos, they are $m_{ee}=m_{e\tau}=0$ and $m_{ee}=m_{e\mu}=0$ in NO, and $m_{\mu\mu}=m_{\tau\tau}=0$ in both NO and IO, which is however disfavored by cosmological data (see also \cite{Priya:2026tpk} for a recent study).

\begin{figure}
    \centering
    \includegraphics[width=\linewidth]{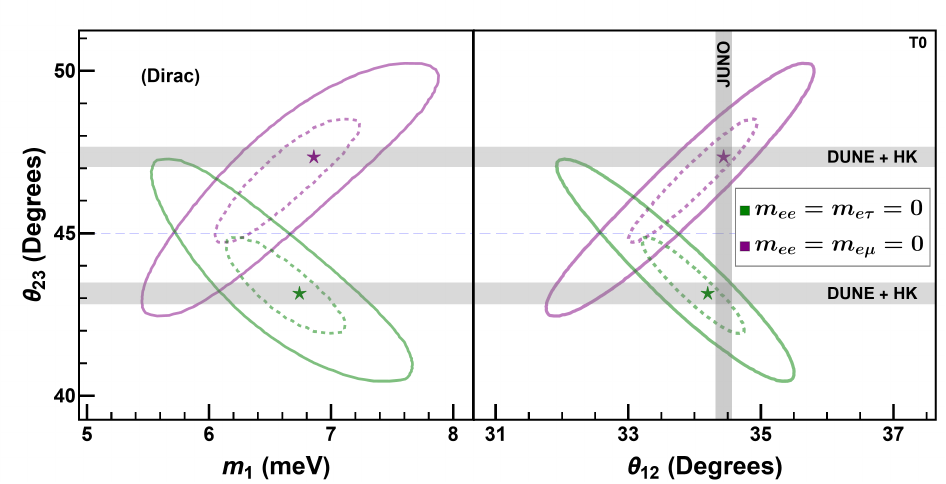}
\caption{
Predictions on $m_1,~\theta_{23},~\theta_{12}$ for the $m_{ee}=m_{e\mu}=0$ and  $m_{ee}=m_{e\tau}=0$ two texture-zeros in the NO for Dirac neutrinos with a  $\Delta\chi^2 = 2.30$ and 11.83 (dashed and solid).  The gray bands refer to the combined $1\sigma$ measurement resolution of DUNE and HK to $\theta_{23}$ at each of the best fit points, assuming reactor constraints on $\sin^2\theta_{13}$ and octant resolution is achieved. 
 JUNO $1\sigma$ sensitivity is based on 6 years of taking data \cite{JUNO:2022mxj} centered at the best fit of $m_{ee}=m_{e\mu}=0$.}
\label{fig:TwoTexZerosContourDirac_Th12}
\end{figure}

Both $m_{ee}=m_{e\tau}=0$, and $m_{ee}=m_{e\mu}=0$ predict a similar, narrow range for $m_1\in [5,9]$ meV; the similarity is due to the approximate $\mu-\tau$ symmetry. The preferred regions are a superposition of the one texture-zero preferred regions.
However, they predict different preferences for the octant of $\theta_{23}$: $m_{ee}=m_{e\tau}=0$ favors the upper octant whereas $m_{ee}=m_{e\mu}=0$ favors the lower octant for $m_1\in[7,9]$ meV, and the preference flips for $m_1\in [5,7]$ meV. Additionally, they predict correlations between $\theta_{23}$ and $\theta_{12}$, and $\theta_{23}$ and $m_1$, see fig.~\ref{fig:TwoTexZerosContourDirac_Th12}.
In these plots, we have assumed the best-fit value of $m_{ee}=m_{e\tau}=0$ to be realized in nature which is measured by DUNE and HK and JUNO with their forecasted sensitivities.  A precise measurement of  $\theta_{12}$ combined with measuring $\theta_{23}$ to be in either octant can distinguish these model predictions. Moreover, these two texture-zeros predict $\delta=0$ or $\pi$ and no preference or correlations for $\theta_{13}$ or the mass splittings.

In NO, $m_{\mu\mu}=m_{\tau\tau}=0$ predicts a narrow range for the lightest mass, $m_1\in [49,52]$~meV, in tension with cosmological data, and makes no prediction on $\cos\delta$. Within this region, it also predicts a very narrow range for $\theta_{23}\in [40.8^\circ , 40.9^\circ]$, making it easily ruled out by a precise measurement of $\theta_{23}$. In IO, $m_{\mu\mu}=m_{\tau\tau}=0$ predicts $m_3\in [23,40]$ meV and no preference for $\theta_{23}$ in its full $3\sigma$ range. These model predictions restrict $\cos\delta\in[-1,0.25]$, and exhibit a strong correlation between $\cos\delta$ and $\theta_{23}$, with $\cos\delta>0$ preferred when $\theta_{23}>41^\circ$, allowing a simultaneous measurement of these two observables to potentially rule out the two texture-zero in IO. 

\subsection{Discussion}

To summarize our discussion of texture-zeros, we find that it is often possible to differentiate both Majorana and Dirac one texture-zeros with next generation data, but the key observable: the octant of $\theta_{23}$, the absolute mass scale, or $\delta$, depends quite subtly on the textures considered and exactly where in parameter space the data lands.
We also see that most of the two texture-zero space is already disfavored due to strong constraints on the mixing parameters and the cosmological neutrino mass scale limit.
The remaining two texture-zeros can be differentiated with precision measurements of the $\theta_{23}$ and the absolute mass scale; for Dirac neutrinos in particular $\theta_{12}$ also plays an important role.

\section{Charged lepton corrections}
\label{sec:clc}
Models with charged lepton corrections assume that the PMNS matrix obtains contributions both from the mixing matrix in the neutrino sector $U_\nu$ and from a non-diagonal mixing matrix in the charged lepton sector $U_e$ as the measurable PMNS matrix is the product of both $U_{PMNS} = U^\dagger_e U_\nu$. The neutrino mixing matrix is assumed to have a form dictated by an underlying discrete symmetry, common choices include $A_4,~S_4, T', A_5$ and others. The discrete symmetry often predicts vanishing $\theta_{13}$, maximal $\theta_{23}$ and specific, model-dependent values for $\theta_{12}$. To account for the experimentally non-vanishing value  of $\theta_{13}$ a 
 non-diagonal charged lepton mixing matrix is required.
Such a non-diagonal charged lepton mixing matrix can arise in models with   grand unified theories based on SU(5) \cite{Georgi:1974sy} or SO(10) \cite{Georgi:1974my,Fritzsch:1974nn} where the structures of the mass matrices for the charged lepton mass matrix and down quarks  coincide \cite{Antusch:2011qg,Marzocca:2011dh,Antusch:2009gu,Antusch:2012fb} leading to a CKM-like mixings in the charged lepton sector   \cite{Datta:2005ci}.

 Models with charged lepton corrections to neutrino mixing matrices predicted by discrete symmetries
 lead to relations between the three mixing angles in the PMNS matrix, $\delta$, and the Majorana phases, which have been well-mapped out analytically in multiple studies  
\cite{Ge:2011ih,Ge:2011qn,Marzocca:2013cr,Petcov:2014laa, Girardi:2014faa, Girardi:2015zva, Girardi:2015vha,Girardi:2016zwz,Girardi:2015rwa,Gehrlein:2022nss} (for reviews, see \cite{King:2013eh,King:2014nza,Petcov:2017ggy}). These model predictions do not involve the neutrino mass scale.

\subsection{Methods and analysis}
For this analysis, we follow the prescription in \cite{Denton:2023hkx}, in which three subclasses of models are considered with a maximum of four rotations split across the neutrino and charged lepton sectors.\footnote{More rotations or different predicted values of the specific mixing angles in the neutrino or charged lepton sectors could arise in certain concrete
models \cite{Frampton:2004ud,deMedeirosVarzielas:2017sdv,Bernigaud:2022sgk}.}
The mixing angles $\theta_{12}^\nu$ and $\theta_{13}^\nu$ are chosen according to popular mixing symmetry forms existing in the literature and $\theta_{23}^\nu$ is always maximal and we study all possible combinations of charged lepton rotation angles. The three subclasses of models are constructed as:
\begin{itemize}
    \item{two rotations in the neutrino sector and one for the charged leptons
    \begin{equation}
        U_{PMNS} = (U_{ij}^e)^\dagger \Psi U_{23}^\nu(\pi/4) U_{12}^\nu(\theta_{12}^{\nu,k})Q,
        \label{eq:CLCTwoNuOneLept}
    \end{equation} with $ij \in \{12,13,23\}$}
    \item{two rotations in the neutrino sector and two for the charged lepton \begin{equation}
        U_{PMNS} = (U_{ij}^e)^\dagger(U_{lm}^e)^\dagger \Psi U_{23}^\nu(\pi/4) U_{12}^\nu(\theta_{12}^{\nu,k})Q,
        \label{eq:CLCTwoTwoLept}
    \end{equation}
    with $ij \in \{12,13\}$, $lm \in \{13,23\}$, $ij\not = lm$}
    \item{three rotations in the neutrino sector and one for the charged lepton \begin{equation}
        U_{PMNS} = (U_{ij}^e)^\dagger\Psi U_{23}^\nu(\pi/4) U_{13}^\nu(\theta_{13}^{\nu,p}) U_{12}^\nu(\theta_{12}^{\nu,k})Q,
        \label{eq:CLCThreeNuOneLept}
    \end{equation} with $ij \in \{12,13, 23\}$}
\end{itemize}
where $k\in \{\text{TBM, BM, GRA, GRB, HG}\}$ represents Tri-Bi-Maximal mixing \cite{Harrison:2002er, Harrison:2002kp,Xing:2002sw,He:2003rm}, Bi-Maximal Mixing \cite{Vissani:1997pa,Barger:1998ta,Baltz:1998ey}, Golden Ratio A \cite{Everett:2008et,Kajiyama:2007gx}, Golden Ratio B \cite{Rodejohann:2008ir,Adulpravitchai:2009bg}, and Hexagonal symmetry \cite{Albright:2010ap,Kim:2010zub} forms of the mixing matrix corresponding to $\sin^2\theta_{12}^{\nu, k} = 1/3, 1/2, 1/(2+\phi), (3-\phi)/4,$ and $1/4$ respectively, with $\phi = (1+\sqrt{5})/2$.  We also consider the additional subclasses with three neutrino rotations where $\theta_{13}^{\nu,p} = \pi/10$, $\pi/20$, and $\arcsin(1/3)$ dubbed T13-1, T13-2, and T13-3, respectfully, motivated by existing models in the literature \cite{Bazzocchi:2011ax,Rodejohann:2014xoa,deAdelhartToorop:2011nfg,Ding:2012xx,King:2012in}.
The $U_{ij}^L$ matrices are rotations in the $(ij)$ sector with mixing angle $\theta_{ij}^L$ for $L\in\{e,\nu\}$.
The phase matrices $\Psi = \diag(1, e^{-i\psi_1}, e^{-i\psi_2})$ and $Q =\diag(1, e^{-i\xi_1/2}, e^{-i\xi_2/2})$ contain the phases in the charged lepton mixing matrix as free parameters. They will affect the prediction on $\cos\delta$.

Within our statistical framework, we report 29 viable charged lepton corrections, listed in table~\ref{tab:CLC} across the 75 possible, with the two listed in red found valid in \cite{Denton:2023hkx} having been ruled out with the inclusion of the recent JUNO data: $[U_{12}^{e}, U_{23}^{e}, \theta_{12}^{\nu,{\rm BM}}]$ predicts $\theta_{12}\sim 36.3^\circ$, and $[U_{12}^{e}, \pi/20, \theta_{12}^{\nu,{\rm GRA}}]$ predicts $\theta_{12}\sim 31.35^\circ$.  See also \cite{Ding:2025dqd,Giarnetti:2026nbd,Petcov:2025aci} for recent studies of charged-lepton corrections in the context of the first data from JUNO. We report 8 valid model predictions with one charged lepton rotation and two neutrino rotations out of the possible 15, 7 valid model predictions containing one charged lepton rotation and three neutrino rotations out of the possible 45, and 14 valid model predictions containing two charged lepton rotations and two neutrino rotations out of the possible 15.

\begin{table}
\centering
\caption{Viable model predictions of charged lepton corrections and the allowed parameter space in $\delta$ and $\theta_{23}$. Expressions with just one charged lepton correction predict $\sin^2\theta_{23}$ within $\pm0.001$. 
The model predictions in red have been ruled out by the recent JUNO measurement of $\theta_{12}$ \cite{JUNO:2025gmd}.}
\begin{tabular}{c|c|c}
Model & $\cos\delta$ & $s_{23}^2$ \\ \hline
$ U_{13}^{e}, \theta_{12}^{\nu,{\rm TBM}} $ & $[-0.30,0.07]$ & $\sim 0.488$ \\
$ U_{13}^{e}, \theta_{12}^{\nu,{\rm GRA}} $ & $[0.12,0.48]$ &$\sim 0.488$ \\
$ U_{13}^{e}, \theta_{12}^{\nu,{\rm GRB}} $ & $[-0.39,-0.02]$ & $\sim 0.488$ \\
$ U_{13}^{e}, \theta_{12}^{\nu,{\rm HEX}} $ & $[0.32,0.66]$ & $\sim 0.488$ \\
$ U_{12}^{e}, \theta_{12}^{\nu,{\rm TBM}} $ & $[-0.07, 0.30]$ & $\sim 0.511$\\
$ U_{12}^{e}, \theta_{12}^{\nu,{\rm GRA}} $ & $[-0.487 0.13]$ & $\sim 0.511$\\
$ U_{12}^{e}, \theta_{12}^{\nu,{\rm GRB}} $ & $[0.02,0.39]$ & $\sim 0.511$ \\
$ U_{12}^{e}, \theta_{12}^{\nu,{\rm HEX}} $ & $[-0.67,-0.32]$ & $\sim 0.511$\\
$ U_{13}^{e}, \pi/20, \theta_{12}^{\nu,{\rm BM}} $ & $[0.03,0.42]$ & $\sim 0.498$  \\
$ U_{13}^{e}, \pi/20, \theta_{12}^{\nu,{\rm GRA}} $ & $[-1,-0.59]$ & $\sim 0.501$ \\
$ U_{13}^{e}, \pi/20, \theta_{12}^{\nu,{\rm HEX}} $ & $[-0.78,-0.38]$ & $\sim 0.501$\\
$ U_{12}^{e}, \pi/20, \theta_{12}^{\nu,{\rm TBM}} $ & $[-1,-0.73]$ & $\sim 0.498$\\
$ U_{12}^{e}, \pi/20, \theta_{12}^{\nu,{\rm GRB}} $ & $[-1, -0.65]$ & $\sim 0.498$\\
$ U_{12}^{e}, \pi/10, \theta_{12}^{\nu,{\rm BM}} $ & $[-0.97,-0.6]$ & $\sim 0.462$ \\
$ U_{12}^{e}, \sin^{-1}\frac13, \theta_{12}^{\nu,{\rm BM}} $ & $[-1,-0.75]$ & $\sim0.454$\\
$ U_{12}^{e}, U_{23}^{e}, \theta_{12}^{\nu,{\rm TBM}} $ & $[-0.45,-0.1]\cup [0.33, 0.65]$ & $[0.41, 0.62]$ \\
$ U_{12}^{e}, U_{23}^{e}, \theta_{12}^{\nu,{\rm GRA}} $ & $[-0.78,-0.511]\cup [-0.08, 0.27]$ & $[0.41,0.62]$\\
$ U_{12}^{e}, U_{23}^{e}, \theta_{12}^{\nu,{\rm GRB}} $ & $[-0.37,-0.01]\cup [0.41,0.72]$ & $[0.41,0.62]$\\
$ U_{12}^{e}, U_{23}^{e}, \theta_{12}^{\nu,{\rm HEX}} $ &$[-0.91, -0.67]\cup [-0.31,0.08]$ & $[0.41, 0.62]$ \\
$ U_{13}^{e}, U_{23}^{e}, \theta_{12}^{\nu,{\rm TBM}} $ & $[-0.38, 0.06]$ & $[0.41, 0.62]$\\
$ U_{13}^{e}, U_{23}^{e}, \theta_{12}^{\nu,{\rm BM}} $ & $[-1,-0.92]$ & $[0.39, 0.47]$ \\
$ U_{13}^{e}, U_{23}^{e}, \theta_{12}^{\nu,{\rm GRA}} $ & $[-0.12,0.54]$ & $[0.41,0.62]$\\
$ U_{13}^{e}, U_{23}^{e}, \theta_{12}^{\nu,{\rm GRB}} $ & $[-0.49, -0.02]$& $[0.41, 0.62]$ \\
$ U_{13}^{e}, U_{23}^{e}, \theta_{12}^{\nu,{\rm HEX}} $ & $[0.32,0.78]$ & $[0.41,0.62]$\\
$ U_{12}^{e}, U_{13}^{e}, \theta_{12}^{\nu,{\rm TBM}} $ & $[-1,1]$ & $[0.49,0.62]$ \\
$ U_{12}^{e}, U_{13}^{e}, \theta_{12}^{\nu,{\rm BM}} $ & $[-1,0.97]$& $[0.50,0.62]$\\
$ U_{12}^{e}, U_{13}^{e}, \theta_{12}^{\nu,{\rm GRA}} $ & $[-0.99,0.99]$ & $[0.49, 0.62]$\\
$ U_{12}^{e}, U_{13}^{e}, \theta_{12}^{\nu,{\rm GRB}} $ & $[-1,1]$ & $[0.49,0.62]$\\
$ U_{12}^{e}, U_{13}^{e}, \theta_{12}^{\nu,{\rm HEX}} $ & $[-0.95,0.96]$& $[0.49,0.62]$\\
\colorbox{LightRed}{$ U_{12}^{e}, U_{23}^{e}, \theta_{12}^{\nu,{\rm BM}} $} & --- & ---\\
\colorbox{LightRed}{$ U_{12}^{e}, \pi/20, \theta_{12}^{\nu,{\rm GRA}} $} & --- & ---\\
\end{tabular}
\label{tab:CLC}
\end{table}

\subsection{Discussion}
\begin{figure}
    \centering
    \includegraphics[width=\linewidth]{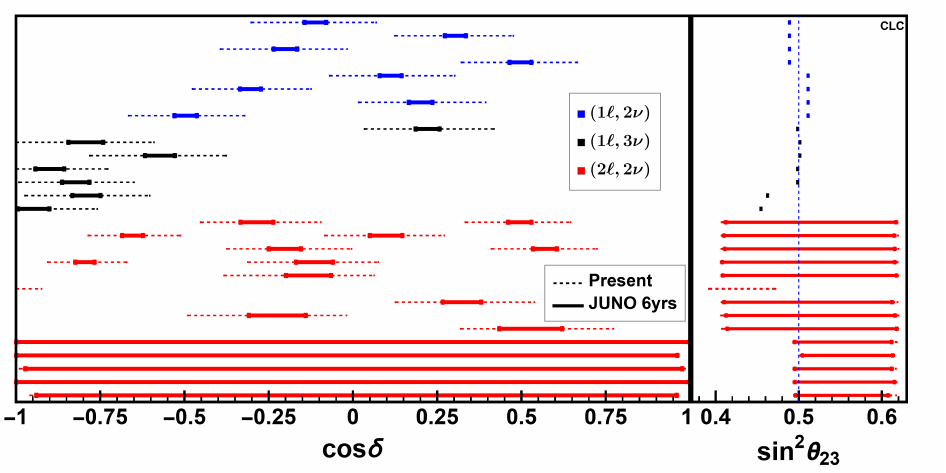}
    \caption{Predictions on $\cos\delta$ and $\sin^2\theta_{23}$ for the 29 valid charged lepton corrections in the order presented in table \ref{tab:CLC}, as classified by the number of rotations in eqs.~(\ref{eq:CLCTwoNuOneLept}-\ref{eq:CLCThreeNuOneLept}).
    Several expressions with two charged lepton rotations and two neutrino rotations predict disjoint regions for $\cos\delta$.
    The dashed bars represent the current model predictions with $\Delta\chi^2<9$, while the solid line shows the predictions assuming JUNO measures the model prediction for $\theta_{12}$ with the expected precision after 6 years \cite{JUNO:2022mxj}. Notably, $ [U_{13}^{e}, U_{23}^{e}, \theta_{12}^{\nu,{\rm BM}}]$  predicts $\theta_{12}$ to be at its current upper experimental $3\sigma$ value. If this model prediction is measured with increased precision, this expression is ruled out and  hence  we only show the dashed line.}
    \label{fig:CLCCosinePredictions}
\end{figure}

We find that all model predictions but one predict $\theta_{12}$ and $\theta_{13}$ to be within their $3\sigma$ bounds, with the exception of $ [U_{13}^{e}, U_{23}^{e}, \theta_{12}^{\nu,{\rm BM}}]$ which predicts $\theta_{12}>34.4^\circ$ and $\cos\delta \sim -1$. Hence, a measurement of $\theta_{12}$ or $\theta_{13}$ alone is unlikely to make substantial progress in ruling out or differentiating between the majority of the model predictions. The predictions for $\sin^2\theta_{23}$ and $\cos\delta$ however, show general trends in the different model subclasses. We show the predicted ranges in table~\ref{tab:CLC} and fig.~\ref{fig:CLCCosinePredictions}. All 15 possible expressions containing only one rotation in the charged lepton sector predict  $\theta_{23}$ within $0.2^\circ$. Moreover, 13 of these predict $\theta_{23}$ to be less than $0.5^\circ$ from maximal, in regions where DUNE and HK would be unable to provide a definitive determination of the octant. However, a measurement of $\theta_{23}$ far from maximal would be able to rule out many of these.  Of the 14 valid expressions with two charged lepton rotations, five predict $\theta_{23}$  mostly in its upper octant, with one, $[ U_{13}^{e}, U_{23}^{e}, \theta_{12}^{\nu,{\rm BM}} ]$, exhibiting a preference for the lower octant, and the remaining eight expressions predict $\theta_{23}$ within its current $3\sigma$ range. 

 The prediction for $\cos\delta$ crucially depends on the predicted value of $\theta_{12}^\nu$ and the measured value of  $\theta_{12}$, 
hence by measuring $\theta_{12}$ more precisely with upcoming JUNO data the prediction on $\cos\delta$ will get more precise. 
Figure \ref{fig:CLCCosinePredictions} shows the predictions on $\cos\delta$ for the valid 29 charged lepton corrections within $\Delta \chi^2<9$ as classified by their rotation structure. The dashed lines represent the current prediction regions  using our statistical analysis, while the solid lines show the reduction of the allowed regions assuming that JUNO measures the model prediction for $\theta_{12}$ with its 6-year precision. Future JUNO data reduces the predicted regions for $\cos\delta$ by a factor of $\sim 3$. Using the current precision on $\theta_{12}$, many model predictions include overlapping regions for $\cos\delta$, making it challenging to differentiate among them with a measurement of  $\cos\delta$ only.  However, general trends are noticeable. Expressions with two rotations in the neutrino sector and one in the charged lepton sector (blue lines) tend to predict near maximal CP-violation, while those with three rotations in the neutrino sector and one in the charged lepton sector (black) tend to predict less CP-violation. Expressions with two rotations in the charged lepton sector (red) tend to produce less constrained predictions, therefore we considered a maximal of four rotations split across both sectors.

To summarize, currently no oscillation measurement alone can easily differentiate among these model predictions, but combined future oscillation data will have a significant impact on our ability to differentiate among them by leveraging the strong correlations between $\theta_{12}$ and $\cos\delta$, as well as information from $\theta_{23}$.

\section{Modular symmetries}
\label{sec:modular}
Another class of highly predictive  flavor models for neutrino  parameters are derived from modular symmetries (MS). These models, originally proposed in \cite{feruglio2017neutrinomassesmodularforms}, see also \cite{Yao:2020zml,Kobayashi:2023zzc,Ding:2023htn}, are derived in supersymmetric theories, where the family symmetry of the Majorana mass matrix is fixed by modular invariance. These models are based on the modular group $SL(2,\mathbb{Z})$, which corresponds to the symmetry group of a torus. The vacuum expectation value of complex modulus $\tau$ acts as the sole source of flavor symmetry breaking. This leads to a reduction of the free parameters in the model, and it has been shown in \cite{Gehrlein:2020jnr} that one can derive constraint equations of the form of eq.~\eqref{eq:GeneralSumRule} for models with a fixed modulus. In these cases the neutrino mass matrix depends on two free model parameters only. These models lead to neutrino mass sum rules of the form in eq.~\eqref{eq:GeneralSumRule} where the coefficients $C_i$ are now functions of the two free model parameters in the neutrino mass matrix which also predict the neutrino mixing parameters. Hence, in addition to a neutrino mass sum rule, we also find relations between the three angles $\theta_{ij}$ and the phase $\delta$ of the PMNS matrix.

\subsection{Method and analysis}
In our analysis, we  study the four model predictions presented in \cite{Gehrlein:2020jnr}, which we call MS 1-4. These are based on a modular $A_4$ symmetry \cite{Novichkov:2018yse}, two modular $S_4$ groups \cite{King:2019vhv}, one modular $S_4$ symmetry \cite{Novichkov:2018ovf}, and one modular $A_5$ symmetry \cite{Novichkov:2018nkm}, respectively.
We also consider two more recent model predictions \cite{deMedeirosVarzielas:2021pug} and \cite{deMedeirosVarzielas:2022ihu} which are based on two modular $A_4$ groups and two modular $A_5$ groups, which we denote MS 5 and MS 6.
\footnote{There are also models with a free value of the modulus field
where a sum rule can arise, like in \cite{CentellesChulia:2023zhu}.}
It is worth noting that MS 5 and MS 6 depend on three free parameters. However, the three parameters themselves are only relevant for the predictions on the Majorana phases, which is not relevant for our analysis, and the remaining mixing parameters can all be re-parametrized in terms of two free variables, as in MS 1-4. 
  
The relations between the neutrino observables  have been worked out in their corresponding works \cite{Gehrlein:2020jnr, Novichkov:2018yse,King:2019vhv,Novichkov:2018ovf,Novichkov:2018nkm,deMedeirosVarzielas:2021pug,deMedeirosVarzielas:2022ihu}. Only the relations between $\theta_{12}$ and $\theta_{13}$ are analytically simple and shown in table \ref{tab:ModularSymRelations}. However, all mixing parameters are correlated, and additionally, since the coefficients of the resulting mass sum rules depend on the mixing parameters, the prediction on the lightest mass also depends on the mixing parameters in a non-trivial way.

\begin{table}
\centering
\caption{The key prediction relating $\theta_{13}$ and $\theta_{12}$ for modular symmetries with a fixed modulus.
We also show the numerical current status of the LHS (depending on $\theta_{12}$) and the RHS (depending on $\theta_{13}$) with $1\sigma$ uncertainties. 
In the final one, $x\equiv\sqrt5\varphi$ where $\varphi\equiv(1+\sqrt5)/2$ is the golden ratio.
See also fig.~\ref{fig:ModularSymNODiscriminatorPlot}.}
\begin{tabular}{c|c|c|c}
Model & Group & Prediction & Current Status\\ \hline
MS 1 \& 5 & $\mathcal{A}_4$ & $s^2_{12} = \frac{1}{3c^2_{13}}$ & $0.309\pm0.009\leftrightarrow0.341\pm0.0002$\\ \hline
MS 2 \& 3 & $\mathcal{S}_4$ &  $t_{12}^2 = \frac{1-3 s_{13}^2}{2}$ & $0.448\pm0.018\leftrightarrow0.467\pm0.0009$\\ \hline
MS 4 \& 6 & $\mathcal{A}_5$ & $t_{12}^2=\frac{1/x}{x/5-s_{13}^2}$ & $0.448\pm0.018\leftrightarrow0.394\pm0.0003$ \\
\end{tabular}
\label{tab:ModularSymRelations}
\end{table}

\begin{figure}
    \centering
    \includegraphics[width=\linewidth]{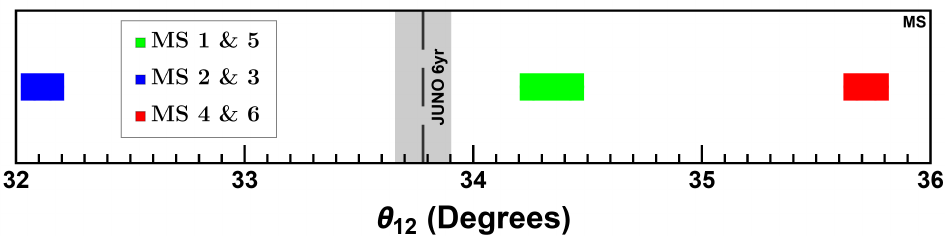}
    \caption{The $\Delta \chi^2 <9$  model predictions on $\theta_{12}$ for modular symmetries. We identify three pairs of modular symmetries which have the same predictions for $\theta_{12}$. The results are the same in both MOs. Also shown is the current JUNO best fit point and its $1\sigma$ sensitivity after 6 years of data taking \cite{JUNO:2025gmd,JUNO:2022mxj}. See also table \ref{tab:ModularSymRelations}.}
    \label{fig:ModularSymTh12}
\end{figure}

 Figure \ref{fig:ModularSymTh12} shows the predicted regions for $\theta_{12}$ for each MS along with the projected sensitivity of JUNO after 6 years of data taking \cite{JUNO:2022mxj}. We find that the predictions for $\theta_{12}$ fall into three groups of MS 1 \& 5, MS 2 \& 3, and MS 4 \& 6 because they are based on the same underlying symmetry.   
The recent JUNO measurements on $\theta_{12}$ are in modest tension with MS 1 \& 5 and MS 4 \& 6 (see \cite{Shang:2026qkh,Gao:2025jlw} for recent studies of different model predictions based on modular symmetries in light of recent JUNO data) and future JUNO data is expected to critically probe these predictions further. In particular, if JUNO measures the current best-fit point of $\theta_{12}$ but with increased precision, all model predictions are in strong tension.

In case the JUNO measurements do not definitively rule out one modular symmetry pair, simultaneous measurements of $\theta_{23}$, $\delta$, and $m_\ell$ can again be used to differentiate these from the remaining pair. Indeed, these model predictions include correlations between $\theta_{23}-\cos\delta$. For further detail, we provide the predictions of the modular symmetries in figs.~\ref{fig:ModularSymDeltaVSTh23}-\ref{fig:ModularSymDeltaVsLogmIO} in the appendix. Figures \ref{fig:ModularSymNODiscriminatorPlot}-\ref{fig:ModularSymIODiscriminatorPlot} show the most important parameters to measure at each given mass scale to distinguish between these MS pairs.  Based on our findings, it may prove difficult to resolve the different predictions between two modular symmetries from measurements of $\theta_{12}$, $\theta_{23}$, and $\delta$ alone, and thus we will require further information on the scale of the lightest mass to make conclusive statements.

\subsection{Discussion}

The predictions of the studied models based on Modular Symmetries are similar to the model predictions of texture-zeros in sec.~\ref{sec:texture zeros}. They make predictions on $\cos\delta$ and predict only lower bounds on the absolute mass scale $m_\ell$. All model predictions predict $\theta_{13}$ and the two mass splittings $\Delta m^2_{ij}$ to be within their $3\sigma$ experimental bounds and exhibit no notable correlations for these parameters.

The relevant predictions made by these pairings in the NO are as follows:

\begin{itemize}
    \item{ Models constructed from a modular $A_4$ symmetry (MS 1 and 5): The predictions of these models vary significantly with the scale of the lightest mass, $m_1$. It is worth noting that MS 1 permits $m_1 = 0$, but requires $\theta_{23}$ to be at the upper end of its permitted $3\sigma$ range at around $\theta_{23}\sim 51^\circ$ for $m_1\lesssim 1$ meV. For $m_1=0$, MS 1 requires
    \begin{equation}
    50.6^\circ \leq \theta_{23} \leq 51.4^\circ \;,\; -0.99 \leq \cos\delta \leq  -0.94 \text{ (MS 1, $m_1=0$).}
    \end{equation}
    The predictions for $\theta_{23}$ and $\delta$ change for $m_1\gtrsim 20$ meV, where MS 1 exhibits a preference for $\theta_{23}\lesssim 48^\circ$ and $\cos\delta>-0.5$. MS 5 permits $m_1 \gtrsim 2$ meV and, when $m_1 \lesssim 6$ meV, predicts $\theta_{23}<45^\circ$ and $\cos\delta >0.$ However, in the range of $ m_1\in [10, 30]$ meV, MS 5 predicts $\theta_{23}>45^\circ$ and $\cos\delta <0$. Then, for $m_1\gtrsim 30$ meV, these predictions expand to $\theta_{23}>43^\circ$ and $\cos\delta<0.5$. Hence, discriminating between these model predictions is highly dependent on a measurement of $m_1$.}
    \item{Models constructed from a modular $S_4$ symmetry (MS 2 and 3):  MS2 predicts $m_1\gtrsim 70$ meV and strictly prefers the upper octant for $\theta_{23}$ and $\cos\delta\in[0,0.54]$. MS3 predicts $m_1\gtrsim 30$ meV, and $\cos\delta\in[-0.42,0.54]$, and no significant prediction on $\theta_{23}$. This means a measurement in the lower octant for $\theta_{23}$ or $\cos\delta<0$ can immediately distinguish these model predictions. These model predictions on $m_1$ also place them in tension with the cosmological bound and will be probed by future measurements of $m_\ell.$}
    \item{Models constructed from a modular $A_5$ symmetry (MS 4 and 6): MS 4 predicts $m_1\gtrsim 4$ meV, but requires $\theta_{23}<43^\circ$ and $\cos\delta > 0.5$ when $m_1\lesssim 20$ meV. MS 6 predicts $m_1\gtrsim 15$ meV, but requires $\theta_{23}>45^\circ$ and $\cos\delta <0$ when $m_1\lesssim 30$ eV. When $m_1\gtrsim 40$ meV, both require $\theta_{23}<48^\circ$, with MS4 also placing a lower bound of $\theta_{23}>41^\circ$ in this same range. For $m_1\gtrsim 60$ meV, MS 4 requires $\cos\delta\in[-0.55,0.70]$, while MS 6 requires $\cos\delta>-0.45$. Discriminating between these model predictions could prove difficult, and will highly depend on the mass scale and the true values of $\theta_{23}$ and $\delta$.}
\end{itemize}
\begin{figure}
    \centering
    \includegraphics[width=\linewidth]{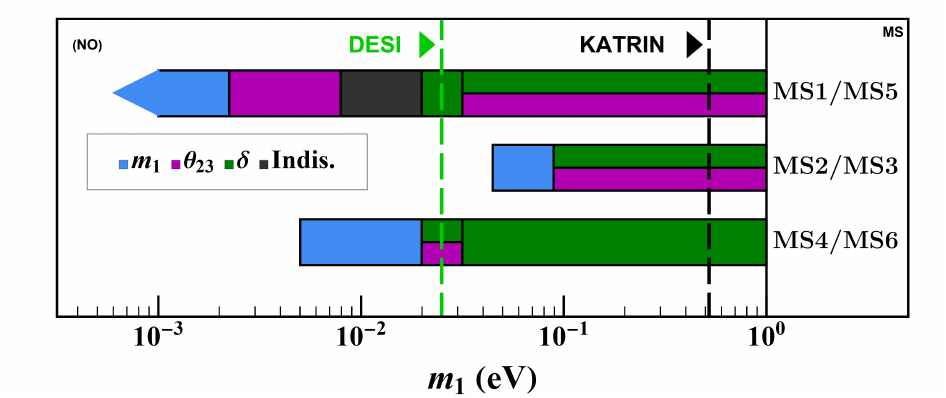}
    \caption{Identification of the most important neutrino  observables to distinguish between modular symmetry pairs in the NO. Indication of $\theta_{23}$ is dependent on the resolution of the octant of $\theta_{23}$. The model predictions are indistinguishable (black) at certain mass scales if they cannot be differentiated using any mixing parameter measurements at that mass scale. Regions with two colors can use either parameter individually to distinguish the two model predictions. Bands with arrows continue on to $m_1 = 0.$ We also show the $3\sigma$ constraints from DESI and KATRIN on the neutrino mass scale.}
    \label{fig:ModularSymNODiscriminatorPlot}
\end{figure}
The relevant predictions made by these pairings of modular symmetries in the IO are as follows:
\begin{figure}
    \centering
    \includegraphics[width=\linewidth]{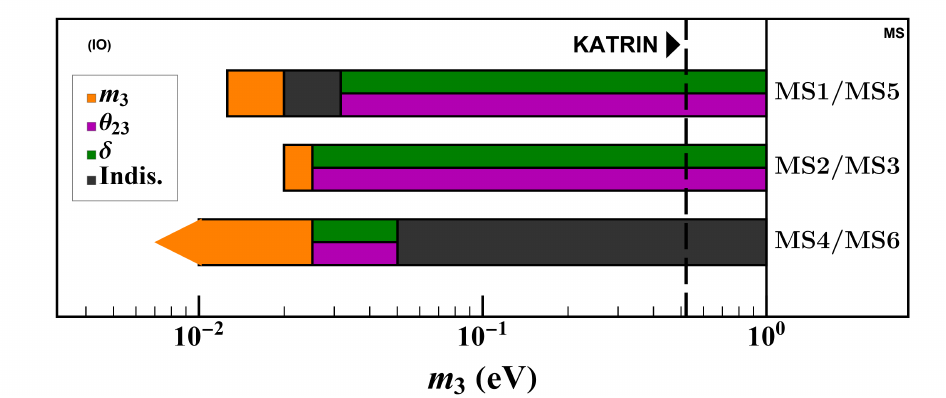}
    \caption{
    Identification of the most important neutrino oscillation observables to distinguish between modular symmetry pairs in the IO. Indication of $\theta_{23}$ is dependent on the resolution of the octant of $\theta_{23}$. The predictions are indistinguishable (black) at certain mass scales if they cannot be differentiated using any mixing parameter measurements at that mass scale. Regions with two colors can use either parameter individually to distinguish the two model predictions. Bands with arrows continue on to $m_3 = 0.$ See fig.~\ref{fig:ModularSymNODiscriminatorPlot} for the results in NO.}
    \label{fig:ModularSymIODiscriminatorPlot}
\end{figure}
\begin{itemize}
    \item{Models constructed from a modular $A_4$ symmetry (MS 1 and 5): MS 1 predicts a lower mass bound of $m_3 \gtrsim 15$ meV, but requires $\theta_{23}<45^\circ$ and $\cos\delta>0$ when $m_3\lesssim 30$ meV. MS 5 predicts $m_3\gtrsim 10$ meV and also requires $\theta_{23}<45^\circ$ when $m_3 \lesssim 30$ meV. This makes the model predictions very difficult to distinguish in the range $m_3\in[15,30]$ meV. However, when $m_3\gtrsim30$ meV, MS 5 predicts $\theta_{23}>43^\circ$ and $\cos\delta >-0.35$, while MS 1 predicts $\theta_{23}<47^\circ$ and $\cos\delta<0.40$, allowing small regions where these model predictions differ. Still, distinguishing these predictions in the case that the true values of the parameters are near $\theta_{23}\sim45^\circ$ and $\cos\delta\sim 0$ could prove difficult.}
    \item{Models constructed from a modular $S_4$ symmetry (MS 2 and 3): MS 2 predicts $m_3\gtrsim 20$ meV and, as in the NO case, requires $\theta_{23}$ to be in its upper octant, as well as requiring $\cos\delta\in[0,0.55]$. MS 3 predicts $m_3 \gtrsim 25$ meV, and $\cos\delta\in[-0.40,0.55]$ and permits $\theta_{23}$ to vary fully within its full $3\sigma$ range, although it has a slight preference for the lower octant at lower masses  around $m_3 \lesssim 30$ meV. Hence, these model predictions can only be distinguished by a measurement of $\theta_{23}$ in its lower octant or $\cos\delta<0$.}
    \item{Models constructed from a modular $A_5$ symmetry (MS 4 and 6): MS 4 does permit $m_3=0$ which requires very narrow ranges for the mixing parameters:
       \begin{equation}
    \theta_{23} \sim 43.8^\circ \;,\; \cos\delta\sim 0.22 \text{ (MS 4, $m_3=0$.)}
    \end{equation}
    When $m_3 \gtrsim 30$ meV, it predicts $\theta_{23}\in[41^\circ,47^\circ]$ and $\cos\delta\in[-0.5,0.68]$.
    MS 6 predicts $m_1\gtrsim 25$ meV. When $m_3\lesssim 50$ meV, MS 6 requires $\theta_{23}$ be in its lower octant and $\cos\delta>0$. For $m_3\gtrsim 50$ meV, $\theta_{23}$ is predicted in its full experimental range, and $\cos\delta >-0.38$. Hence, at higher mass scales, these model predictions disagree very minimally and it could prove difficult to differentiate them unless $\theta_{23}$ is far from maximal or $\cos\delta \sim 1$.}
\end{itemize}
To summarize our discussion of models based on modular symmetries, we conclude that increasing precision on  $\theta_{12}$ is by far the most promising way to rule out these models. Indeed, as these model predictions contain strong correlations between $\theta_{12}$ and $\theta_{13}$, the expected precision of JUNO is enough to confidently rule out all six depending on the true value of $\theta_{12}$. 
In the case that JUNO achieves its precision and all six are not ruled out, only one pair can remain viable at once, and we have shown that the majority of these pairs can be differentiated using combined measurements on the parameters $m_\ell$, $\theta_{23}$, and $\cos\delta$, as the exact predictions of these models is highly dependent on the mass scale.

\section{Constrained sequential dominance}
\label{sec:sequential}
Finally, we study a class of flavor models  based on constrained sequential dominance (CSD)  \cite{King:1998jw,King:1999cm}. This scheme admits an effective two right-handed neutrino model in which the lightest neutrino is massless, $m_1 = 0$, and hence is only valid in the NO \cite{King:2005bj, Antusch:2011ic}. 

\subsection{Methods and analysis}
The general CSD$(n)$ scheme fixes the two columns of the Dirac neutrino matrix to be proportional to $(0,1,-1)$ and $(1,n,2-n)$ in the basis where the right-handed neutrino mass matrix is diagonal, leading to a general mass matrix for light neutrinos of the form
$$M_\nu = a\begin{pmatrix}
    0 & 0&0\\
    0&1&-1\\
    0&-1&1
\end{pmatrix}+ b e^{i\eta}\begin{pmatrix}
    1 & n&2-n\\
    n&n^2&n(2-n)\\
    2-n&n(2-n)&(2-n)^2
\end{pmatrix}$$
where $a,b$ are real model parameters, and $\eta$ is a free phase. For a given choice of $n$, a constrained sequential dominance model takes two free parameters,  the ratio $a/b$ and the phase $\eta$, to make  predictions on all neutrino mixing parameters. It is worth noting that some models propose fixed values of $\eta$ based on an additional $\mathbb{Z}_3$ symmetry, which fixes $e^{i\eta}$ to a cube root of unity \cite{King:2015dvf, Ding:2013hpa,Feruglio:2013hia}, but in the interest of generality, we permit it here to vary freely.
Analytical expressions for the observables from these models have been derived in \cite{deMedeirosVarzielas:2022fbw,King:2015dvf}. Again, these model predictions are always predictive on $\cos\delta$.

We consider five CSD($n$) models studied in the literature that are phenomenologically viable, namely CSD$(3)$ \cite{King:2013iva,Ballett:2016yod}, CSD$(4)$ \cite{King:2013xba}, CSD$(-1/2)$ \cite{Chen:2019oey}, and CSD$(1\pm \sqrt6)$ \cite{deMedeirosVarzielas:2022fbw}.

\begin{figure}
    \centering
    \includegraphics[width=.89\linewidth]{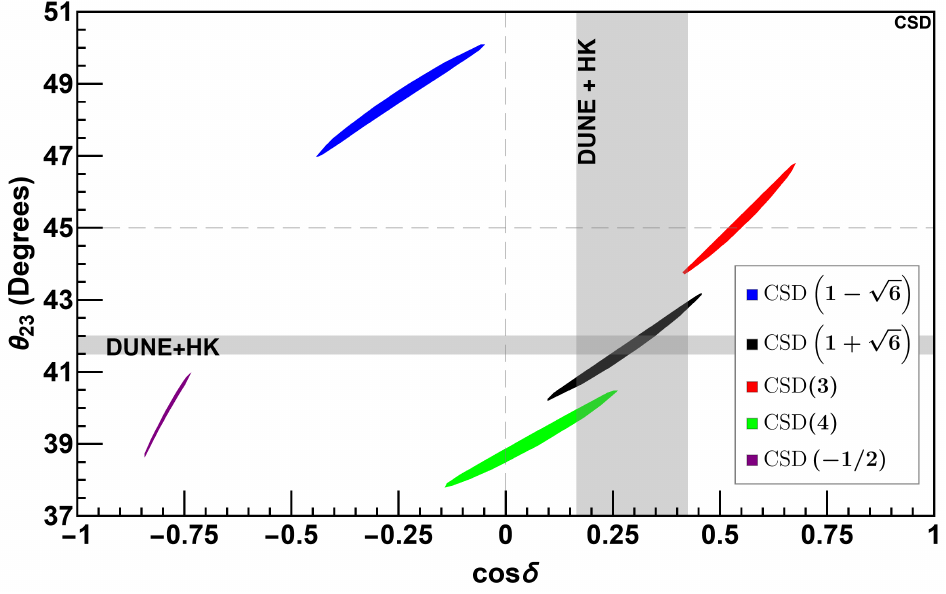}
    \caption{Model prediction parameter spaces of the five CSD$(n)$ with $\Delta\chi^2 < 11.83$ in the $\theta_{23}-\cos\delta$ plane. In gray, we show the combined $1\sigma$ sensitivities from DUNE+HK  assuming they measure  the best fit point of CSD$(1+\sqrt{6})$ (black) for both mixing parameters.}
    \label{fig:LittlestSeeSawContours}
\end{figure}

\subsection{Discussion}
These model predictions contain very strong correlations between $\theta_{12}$ and $\theta_{13}$, with the same relation as for MS 2 and MS 3, see table \ref{tab:ModularSymRelations} and fig.~\ref{fig:ModularSymTh12}. We find that all five prefer $\theta_{12}$ close to 34.3$^\circ$, at the upper $1\sigma$ limit from JUNO, while all predict $\theta_{13}$ and the mass splittings within their experimental $3\sigma$ ranges.
The model predictions also include strong correlations between $\theta_{23}$ and $\delta$.
We present each of the five model prediction parameter spaces in the $\theta_{23}$-$\cos\delta$ plane in fig.~\ref{fig:LittlestSeeSawContours}. The predictions of these five models do not overlap for $\Delta\chi^2<11.83$, and are for the most part easily distinguished with a measurement of either parameter. For those spaces closer to overlapping, we show the combined sensitivity of DUNE and HK at the best fit point of CSD($1+\sqrt{6})$ \cite{DUNE:2024wvj}. We find that a simultaneous measurement of $\theta_{23}$ and $\delta$ from DUNE alone is enough to effectively distinguish between these model predictions, and we expect to be able to rule out many of these in the near future.

A comprehensive review of these models in light of the recent JUNO results has been conducted in \cite{Shang:2026qkh}. Here, we  emphasize the ability of upcoming neutrino experiments at DUNE and HK to further probe these model predictions.

\section{Discussion}
\label{sec:discussion}
While the preceding sections of the text have shown how a given flavor model prediction is impacted by various neutrino observables, in this section, we summarize the results in the opposite order and show which measurements can distinguish between flavor models.
The key conclusions of this section are highlighted in table \ref{tab:ModelComparison}.

\paragraph{$\boldsymbol{m_\ell}$:}
A measurement of the absolute mass scale has implications for many classes of models.
In particular, a determination of a nonzero value of the mass scale from a non-oscillation experiment, either beta decay endpoint or cosmology, would have a significant impact across many flavor model predictions.
In fact, with the exception of charged lepton corrections, significant power to disfavor and discriminate model predictions will be achieved  as the precision on the absolute mass scale is nailed down.
Some key thresholds exist; one important example is $m_1\sim6$ meV in the NO which significantly enhances model prediction discrimination and sensitivity for texture-zeros, and will be achievable by upcoming cosmology measurements.

Future precision on this parameter can be enough to make statements on the validity and difference between models, but it will be challenging.
Specifically, according to \cite{DESI:2016fyo} (see also \cite{Font-Ribera:2013rwa}) DESI combined with Planck and Lyman-$\alpha$ may achieve a precision on the sum of neutrino masses of 0.02 eV.
Then, one will be able to disfavor $m_1=1$ (8) meV at $2\sigma$ when $m_1>19$ (25) meV, which corresponds to $\sum m_i>93$ (107) meV, which is just about at the current limit.
So the neutrino mass scale will need to be near the current limit in order to disfavor key mass ranges.

\paragraph{Atmospheric mass ordering:}
The sign of $\Delta m^2_{31}$, the mass ordering, will be known at high significance in the coming years.
The impact on flavor models will be significant as the predictions and validity are often completely different in different mass orderings.
Only charged-lepton corrections are unaffected by the mass ordering.

\paragraph{$\boldsymbol{\theta_{12}}$:}
The solar mixing angle, now best measured by reactor neutrinos, is a key discriminator for many classes of models.
We find that, especially with the latest JUNO results, a number of specific model predictions that were considered viable are now in increased tension with the data, specifically in the modular symmetries and two texture-zeros, especially when combined with an octant determination for the latter model class. The impact on mass sum rules and constrained sequential dominance, however, is modest to small.

\paragraph{$\boldsymbol{\theta_{23}}$:}
A measurement of the octant of $\theta_{23}$ can be used to significantly cut down the model prediction parameter space for many classes of models with the exception of mass sum rules.
For many pairs of texture-zeros it is the key differentiation parameter, especially in the regions where the absolute mass scale is easiest to probe.
In addition, while the octant is usually the relevant observable, we see for many of the charged lepton correction predictions that the only remaining viable parameter space has a very specific prediction for $\theta_{23}$, often fairly close to 45$^\circ$.
For some of them, we see that the predictions are within 0.5$^\circ$ of maximal which means that any determination of non-maximal mixing would disfavor these scenarios (and vice-versa), something that can happen from $\nu_\mu$ disappearance alone without an input from the more challenging $\nu_\mu\to\nu_e$ appearance channel.
\begin{table}
  \centering
  \caption{Overview of the most discriminating neutrino parameters to  constrain and differentiate between model predictions. The mixing parameters $\Delta m_{ij}^2$ and $\theta_{13}$ are omitted due to their already high precision and minor impact within the five chosen model classes.}
  \begin{threeparttable}
    \renewcommand{\arraystretch}{1.5}
    \begin{tabular}{l | p{9cm} | c c c c c}
    \toprule
    \toprule
  \multicolumn{2}{l}{Observable}  & SR & T0 & CLC & MS & CSD \\
    \midrule
    $m_\ell$ & Measuring $\lesssim 10$ meV will exclude many model predictions; a precise measurement will differentiate many.    
    & $\bigstar$ & $\checkmark$ & $\times$ & $\checkmark$ & $\times$ \\
        \hline
    MO & Resolving the ordering will instantly rule out a significant proportion of all model predictions. & $\checkmark$ & $\checkmark$& $\times$ & $\checkmark$ & $\bigstar$ \\
    \hline
    $\theta_{12}$ & Increased precision alone will rule out some model predictions and will tighten parameter space for others. & $\times$ & $\checkmark$ & $\checkmark$ & $\bigstar$ & $\bigstar$ \\
    \hline
    $\theta_{23}$ & Resolving octant degeneracy alone will have a large impact on most model predictions. & $\times$ & $\checkmark$ & $\checkmark$ & $\checkmark$ & $\checkmark$ \\
    \hline
    $\cos\delta$ & Impact will depend on the true value, but can be leveraged with the other mixing parameters. & $\times$ & $\checkmark$ & $\checkmark$ & $\checkmark$ & $\checkmark$ \\
    \midrule
    \multicolumn{7}{l}{$\times$: Increased precision will not have significant impact.} \\
    \multicolumn{7}{l}{$\checkmark$: Increased precision and synergistic measurements will have notable impact.} \\
    \multicolumn{7}{l}{$\bigstar$: Increased precision alone will have significant impact.} \\
    \bottomrule
    \bottomrule
    \end{tabular}
  \end{threeparttable}
  \label{tab:ModelComparison}
\end{table}

\paragraph{$\boldsymbol{\cos\delta}$:}
The complex phase, being largely undetermined by current oscillation data, is a key discriminator in many cases, with the exception of mass sum rules, meaning a precision measurement of $\delta$ will have a significant impact on our understanding of flavor models.
This parameter is often correlated with $\theta_{23}$.
We emphasize that a measurement of  $\cos\delta$ rather than $\delta$, $\sin\delta$, or the Jarlskog invariant $J$ \cite{Jarlskog:1985ht} is required, as flavor model predictions almost always depend on $\cos\delta$, not one of the other variations.
As the appearance $\nu_\mu\to\nu_e$ channel is most sensitive to $\sin\delta$, is  important to have other probes, such as $\nu_\mu\to\nu_\mu$ which is sensitive to $\cos\delta$ \cite{Denton:2023qmd}, to ensure that the sign degeneracy is broken and the maximum precision possible is achieved.

\paragraph{$\boldsymbol{\Delta m^2_{ij}}$:}
We have not directly addressed the role of improved measurements of $\Delta m^2_{31}$ and $\Delta m^2_{21}$ in the coming years.
We have performed careful checks and we find that for model classes which feature a mass sum rule of the form eq.~\eqref{eq:GeneralSumRule} the mass splittings do not play a role. This is also true of the constrained sequential dominance model class in sec.~\ref{sec:sequential}, which predicts the ratio of the mass splittings fully within the range permitted by experiment. We find that the freedom in other parameters can accommodate for these mass splittings across their experimental ranges, and increased precision on these parameters will ultimately not serve to significantly help us rule out or discriminate any of the model predictions presented in this work.

\paragraph{$\boldsymbol{\theta_{13}}$:}
The high precision of this parameter means that, when it is relevant for a model prediction, its uncertainty is much smaller than the others in the problem.
Thus the modest expected improved precision in the global picture with the addition of the final long-baseline measurements from DUNE and HK beyond the current data from Daya Bay and RENO will not have a significant effect on the flavor model picture.

\section{Summary and conclusions}
\label{sec:conclusions}
In this manuscript, we have  presented and contrasted the predictions from five widely studied classes of flavor model predictions. We have focused on models which, in addition to predicting the parameters in the neutrino sector, also exhibit correlations between them. Namely, we have analyzed models with mass sum rules which predict relations among the light neutrino masses leading to correlated predictions for the lightest mass, the Majorana phases, and the mass splittings, models based on texture-zeros which feature one or two zero entries of the complex Majorana or Dirac mass matrix, leading to correlations between the lightest neutrino mass, and all mixing parameters, models based on charged lepton corrections which feature a non-diagonal charged lepton mixing matrix leading to relations between the measurable mixing angles and phase,  models based on  modular symmetries with a fixed modulus, and models with constrained sequential dominance make correlated predictions on the lightest mass (predicted to be zero in constrained sequential dominance models) and the mixing parameters.
See fig.~\ref{fig:Model_Overview} for an overview of the predictions made by each model class.

We find that out of the 152 model predictions we have studied, 73 are valid within the current experimental $3\sigma$ ranges of the measured neutrino oscillation parameters, while some model predictions include a lightest mass prediction in tension with cosmological results.
With current experimental knowledge on the neutrino parameters the model prediction parameter spaces overlap considerably, this is especially true for mass sum rules, texture-zeros, and  charged lepton corrections model predictions. The model prediction parameter spaces for modular symmetries with a fixed modulus and constrained sequential dominance, however, are much more  distinct. 
Overall, the most promising neutrino parameters to constrain flavor models are the absolute neutrino mass scale, $\theta_{12},\theta_{23}$, $\cos\delta$, and the neutrino mass ordering. Future neutrino experiments are expected to make significant progress on measuring these parameters with increased precision which will lead to improved constraints and distinction between the model predictions.

On the other hand, the mass splittings and $\theta_{13}$ play a sub-dominant role in probing flavor models either because flavor models do not make concrete predictions on them, as is the case for the mass splittings, or because $\theta_{13}$ is already precisely measured, constraining the model predictions on this parameter to an already very narrow region to be valid. 

The neutrino sector is just one part of the
flavor sector of the Standard Model, however 
it plays a significant role in exploring the potential underlying theory of flavor due to the large mixings and small masses in the neutrino sector. Here we have shown that expected results from future neutrino experiments will provide a major impact on the validity and distinction of flavor models in the neutrino sector and as such they play a valuable role  in our
quest for the theory of flavor.

\section*{Acknowledgments}
The work of P.B.D.~is supported by the US Department of Energy under Grant Contract DE-SC0012704. J.G.~and H.T.~were partially supported by the U.~S.~Department of Energy Office of Science under award number DE-SC0025448.

\appendix
\section{Additional correlations for modular symmetries with a fixed modulus}
\begin{figure}
    \centering
    \includegraphics[width=.82\linewidth]{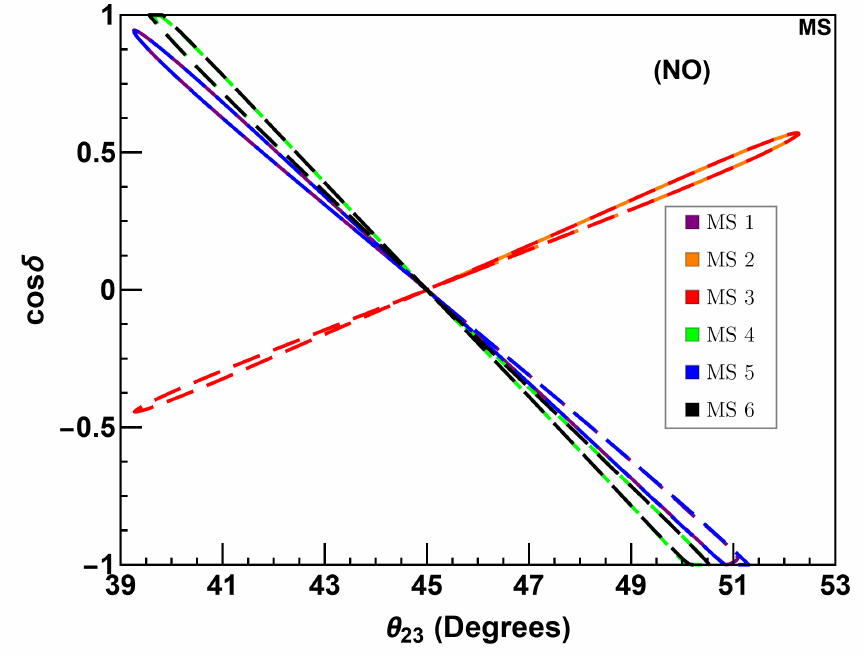}
    \caption{Predictions of the modular symmetries for the NO in the $\theta_{23}$-$\cos\delta$ plane for $\Delta \chi^2 <11.83$.}
    \label{fig:ModularSymDeltaVSTh23}
\end{figure}

\begin{figure}
    \centering
    \includegraphics[width=0.82\linewidth]{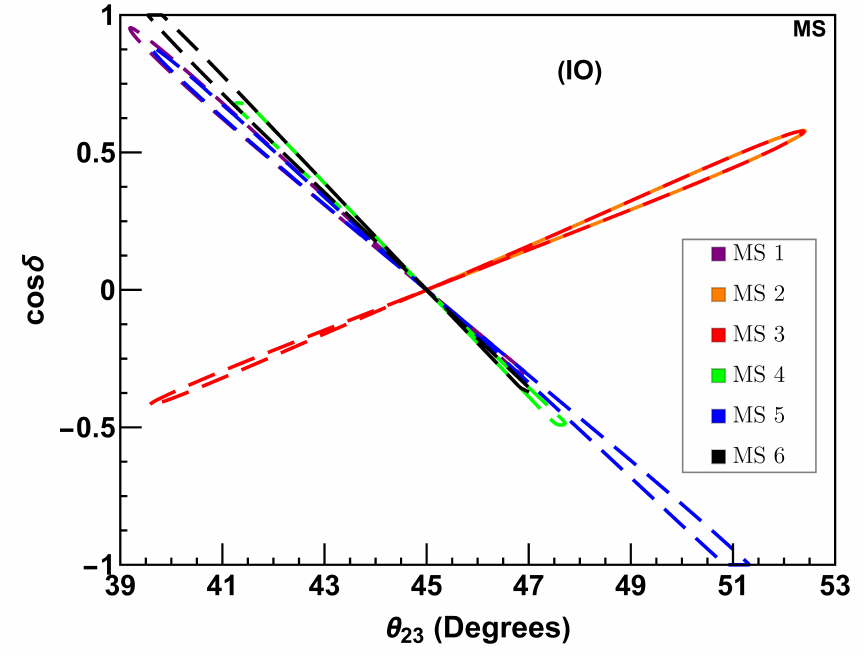}
    \caption{Predictions of the modular symmetries for the IO in the $\theta_{23}$-$\cos\delta$ plane for $\Delta \chi^2 <11.83$.}
    \label{fig:ModularSymDeltaVSTh23IO}
\end{figure}

\begin{figure}
    \centering
    \includegraphics[width=.82\linewidth]{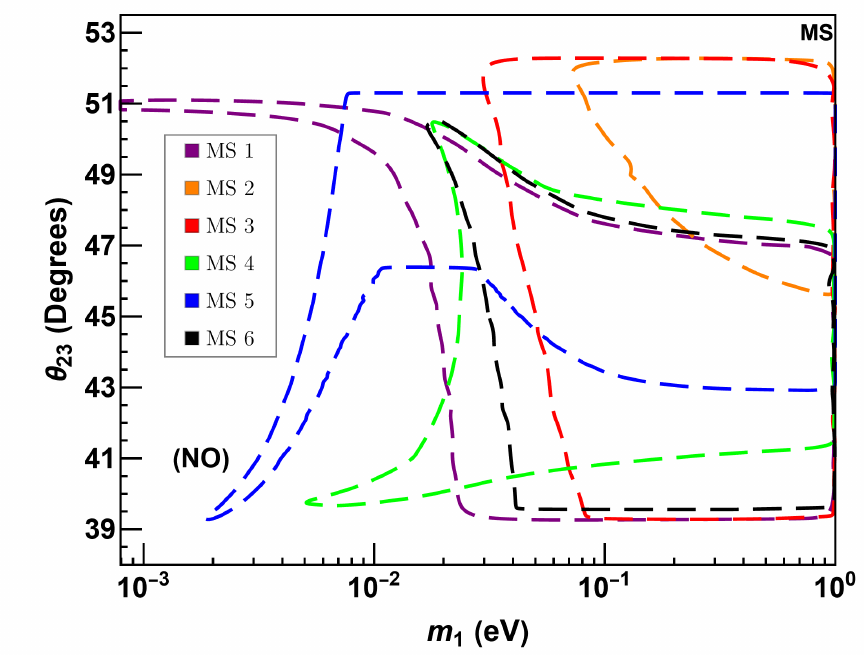}
    \caption{Predictions of the modular symmetries for the NO in the $m_1$-$\theta_{23}$ plane for $\Delta \chi^2 <11.83$.}
    \label{fig:ModularSymTh23VsLogm}
\end{figure}

\begin{figure}
    \centering
    \includegraphics[width=0.82\linewidth]{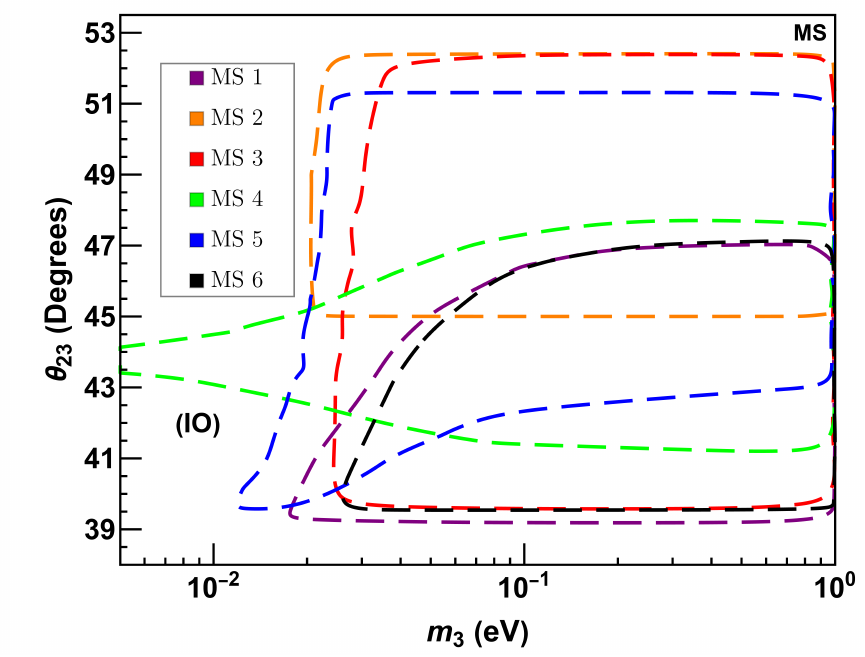}
    \caption{Predictions of the modular symmetries for the IO in the $m_3$-$\theta_{23}$ plane for $\Delta \chi^2 <11.83$.}
    \label{fig:ModularSymTh23VsLogmIO}
\end{figure}

\begin{figure}
    \centering 
    \includegraphics[width=0.82\linewidth]{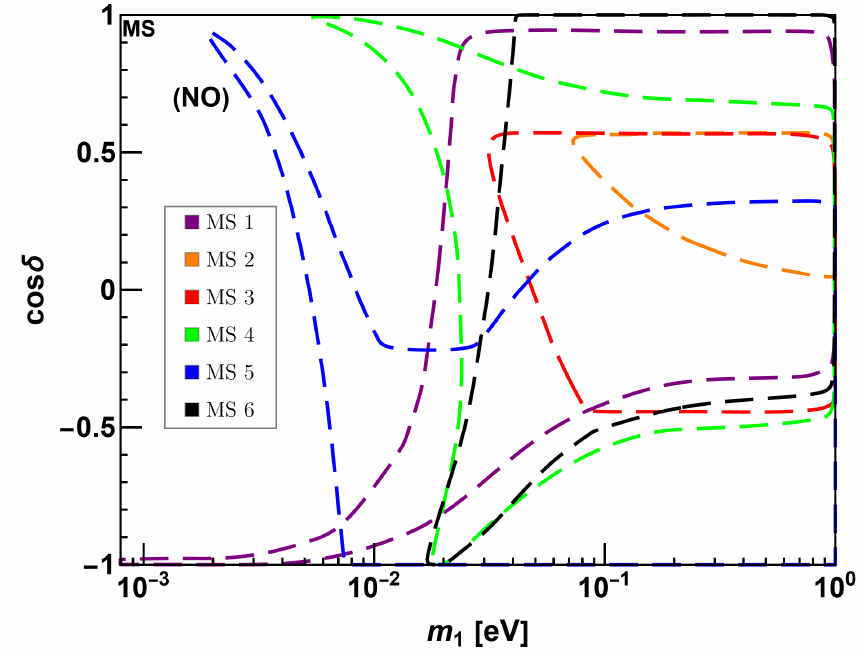}
    \caption{Predictions of the modular symmetries for the NO in the $m_1$-$\cos\delta$ plane for $\Delta \chi^2 <11.83$.}
    \label{fig:ModularSymDeltaVsLogm}
\end{figure}

\begin{figure}
    \centering
    \includegraphics[width=0.82\linewidth]{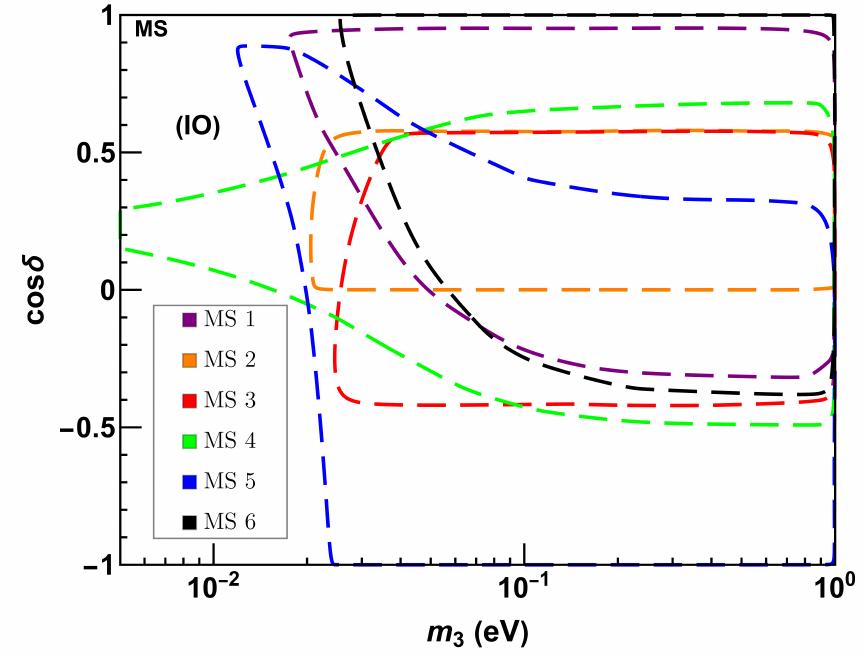}
    \caption{Predictions of the modular symmetries for the IO in the $m_3$-$\cos\delta$ plane for $\Delta \chi^2 <11.83$.}
    \label{fig:ModularSymDeltaVsLogmIO}
\end{figure}

In figs.~\ref{fig:ModularSymDeltaVSTh23}, \ref{fig:ModularSymTh23VsLogm}, \ref{fig:ModularSymDeltaVsLogm} we show additional model prediction parameter spaces and their correlations between $\theta_{23}$, $\cos\delta$ and the lightest mass and $\theta_{23}-\cos\delta$ in the NO for the six modular symmetries with a fixed modulus studied in the main text. The corresponding plots in IO are figs.~\ref{fig:ModularSymDeltaVSTh23IO}, \ref{fig:ModularSymTh23VsLogmIO}, \ref{fig:ModularSymDeltaVsLogmIO}.

\newpage
\bibliographystyle{JHEP}
\bibliography{main}

\end{document}